\documentclass[12pt]{article}
\usepackage{amsmath}
\usepackage{amstext}
\usepackage{amsthm}
\usepackage{amssymb}
\usepackage{mathrsfs} 
\usepackage{dsfont} 
\usepackage[bottom]{footmisc}
\usepackage{indentfirst}
\usepackage{upref}
\usepackage{cite}
\usepackage{units}
\usepackage{graphicx}
\usepackage[nooneline,centerlast]{caption2}
\usepackage{fancybox}
\usepackage[lflt]{floatflt}
\renewcommand{\baselinestretch}{1.30}
\begin{document}
\begin{titlepage} 
\renewcommand{\baselinestretch}{1.3}
\small\normalsize
\begin{flushright}
hep-th/0410119\\
MZ-TH/04-15
\end{flushright}

\vspace{0.1cm}

\begin{center}   

{\LARGE \textsc{Quantum Gravity\\[2.35mm] at Astrophysical Distances?}}

\vspace{1.4cm}
{\large M.~Reuter and H.~Weyer}\\

\vspace{0.7cm}
\noindent
\textit{Institute of Physics, University of Mainz\\
Staudingerweg 7, D--55099 Mainz, Germany}\\

\end{center}   

\vspace*{0.6cm}
\begin{abstract}
  Assuming that Quantum Einstein Gravity (QEG) is the correct theory
  of gravity on all length scales we use analytical results from
  nonperturbative renormalization group (RG) equations as well as
  experimental input in order to characterize the special RG
  trajectory of QEG which is realized in Nature and to determine its
  parameters. On this trajectory, we identify a regime of scales where
  gravitational physics is well described by classical General
  Relativity. Strong renormalization effects occur at both larger and
  smaller momentum scales. The latter lead to a growth of Newton's
  constant at large distances. We argue that this effect becomes
  visible at the scale of galaxies and could provide a solution to the
  astrophysical missing mass problem which does not require any dark
  matter. We show that an extremely weak power law running of Newton's
  constant leads to flat galaxy rotation curves similar to those
  observed in Nature. Furthermore, a possible resolution of the
  cosmological constant problem is proposed by noting that all RG
  trajectories admitting a long classical regime automatically give
  rise to a small cosmological constant.
\end{abstract}
\end{titlepage}
%
%
%
%
%
%
\section{Introduction}
\label{s1}
During the past few years, in the light of a series of investigations
\cite{mr,percadou,oliver1,frank1,oliver2,souma,percacciperini,frank2,litimgrav},
it appeared increasingly likely that Quantum Einstein Gravity (QEG),
the quantum field theory of gravity whose underlying degrees of
freedom are those of the spacetime metric, can be defined
nonperturbatively as a fundamental, ``asymptotically safe''
\cite{wein} theory. By definition, its bare action is given by a
non--Gaussian renormalization group (RG) fixed point. In the framework
of the effective average action \cite{avact,avactrev,ym} a suitable
fixed point is known to exist in the Einstein--Hilbert truncation of
theory space \cite{mr,souma,oliver1} and a higher--derivative
generalization \cite{oliver2} thereof. Detailed analyses of the
reliability of this approximation \cite{oliver1,frank1,oliver2} and a
conceptually independent investigation \cite{max} suggest that the
fixed point should indeed exist in the exact theory, implying its
nonperturbative renormalizability.

The general picture regarding the RG behavior of QEG as it has emerged
so far points towards a certain analogy between QEG and non--Abelian
Yang--Mills theories, Quantum Chromo--Dynamics (QCD) say. For example,
like the Yang--Mills coupling constant, the running Newton constant
$G=G(k)$ is an asymptotically free coupling, it vanishes in the
ultraviolet (UV), i.\,e.\ when the typical momentum scale $k$ becomes
large. In QCD the realm of asymptotic freedom, probed in deep
inelastic scattering processes, for instance, is realized for momenta
$k$ larger than the mass scale $\Lambda_{\text{QCD}}$ which is
induced dynamically by dimensional transmutation. In QEG the analogous
role is played by the Planck mass $m_{\text{Pl}}$. It delimits the
asymptotic scaling region towards the infrared (IR). For $k \gg
m_{\text{Pl}}$ the RG flow is well described by its linearization
about the non--Gaussian fixed point \cite{frank1}.  Both in QCD and QEG
simple local truncations of the running Wilsonian action (effective
average action) are sufficient above $\Lambda_{\text{QCD}}$ and
$m_{\text{Pl}}$, respectively. However, as the scale $k$ approaches
$\Lambda_{\text{QCD}}$ or $m_{\text{Pl}}$ from above, many
complicated, typically nonlocal terms are generated in the effective
action \cite{gluco,frank2,oliver0}. In fact, in the IR, strong
renormalization effects are to be expected because gauge
(diffeomorphism) invariance leads to a massless excitation, the gluon
(graviton), implying potential IR divergences which the RG flow must
cure in a dynamical way. Because of the enormous algebraic complexity
of the corresponding flow equations it is extremely difficult to
explore the RG flow of QCD or QEG in the IR, far below the UV scaling
regime, by purely analytical methods. In QCD lattice techniques can be
used to study the IR sector, but despite recent progress on dynamical
triangulations \cite{amb,ajl} there exists no comparable tool for
gravity yet.

In QCD we have another source of information about its small momentum
or large distance regime. If we take it for granted that QCD is the
correct theory we can exploit the available experimental data on the
strong interaction, interpret them within this theory, and thus obtain
information about the quantum dynamics of QCD, in particular its
nonperturbative IR sector, from the purely phenomenological side. An
example to which we shall come back in a moment are the
non--relativistic quark--antiquark potentials extracted from
quarkonium data (and confirmed on the lattice). They suggest that
nonperturbative IR effects modify the classical Coulomb term by adding
a confinement potential to it which increases (linearly) with
distance:
\begin{align}
V (r) = - \frac{a}{r} + \kappa \,r.
\label{1}
\end{align}
Here $a$ and the string tension $\kappa$ are constants \cite{joos}.

In this paper we are going to apply a similar ``phenomenological''
strategy to gravity. Under the assumption that QEG is the correct
theory of gravity on all distance scales, we try to describe and
characterize the distinguished RG trajectory which is realized in
Nature as completely as possible. We use both observational input and
the available analytical RG studies.

We shall start from the flow equations in the Einstein--Hilbert
truncation, determine which type of its RG trajectories the one
realized in Nature belongs to, identify a regime on it where standard
General Relativity is valid, and finally argue that this regime does
not extend to arbitrarily large distances. In fact, for $k$ smaller
than the momenta typical of the regime of standard gravity, the
truncation predicts a strong increase of $G(k)$ with decreasing $k$,
which, at a certain critical value of $k$, becomes infinite even. The
diverging behavior is clearly an artifact of an insufficient
truncation, but we shall see that the growth of Newton's constant with
the distance can be understood on general grounds as due to a
potential IR singularity. It is the main hypothesis of the present
paper that a ``tamed'' form of this nonperturbative IR growth of
$G(k)$ is a genuine feature of exact QEG.\footnote{Using different
  methods or models, IR quantum gravity effects have also been studied
  in refs.\ \cite{tsamis,mottola,bertodm,bertoproc}.}

\vspace*{2\baselinestretch pt} The problem of the missing mass or
``dark matter'' is one of the most puzzling mysteries of modern
astrophysics and cosmology \cite{padman}.  It has been known for a
long time that the luminous matter contained in a galaxy, for
instance, does not provide enough mass to explain the gravitational
pull the galaxy exerts on ``test masses'' in its vicinity. Typically
their rotation curves $v(r)$, the orbital velocity as a function of
the distance, are almost flat at large distances rather than fall off
according to Kepler's law \cite{combbook}.  Similar, but even stronger
mass discrepancies are observed on all larger distance scales and in
particular in cosmology. The recent high--redshift supernova and CMBR
data show very impressively that the known forms of baryonic matter
account only for a small percentage of the matter in the Universe. A
possible way out is the assumption that the missing mass is due to
some sort of ``dark matter'' which would manifest itself only by its
gravitational effects. However, as to yet it has not been possible to
convincingly identify any dark matter candidate, and so it might be
worthwhile to think about alternatives.

It is a very intriguing idea that the apparent mass discrepancy is not
due to an unknown form of matter we have not discovered yet but rather
indicates that we are using the wrong theory of gravity, Newton's law
in the non--relativistic and General Relativity in the relativistic
case. In fact, Milgrom \cite{mond} has developed a phenomenologically
very successful non--relativistic theory, called MOdified Newtonian
Dynamics or ``MOND'', which explains many properties of galaxies, in
particular their rotation curves, in a unified way without invoking any
dark matter. In its version where gravity (rather than inertia) is
modified, a point mass $M$ produces the potential
\begin{align}
\phi (r) = 
- \frac{\overline{G} M}{r} 
+ \sqrt{a_{0} \, \overline{G} M \,} \, \ln (r)
\label{2}
\end{align}
where $\overline{G}$ and $a_{0}$ are constants. The second term on the
RHS of \eqref{2} is responsible for the flat, non--Keplerian rotation
curves at large distances. So far no wholly satisfactory relativistic
extension of MOND is known.

Also the relativistic theory proposed by Mannheim \cite{mannheim}
where the Lagrangian is the square of the Weyl tensor tries to explain
the rotation curves as due to a non--Newtonian force. The corresponding
potential is of the form
\begin{align}
\phi (r) =
- \frac{\overline{G} M}{r}
+ \widetilde \kappa \, r.
\label{3}
\end{align}
The resulting rotation curves do not become flat but still seem to be
in accord with the observations.

For a detailed discussion of other attempts at modifying gravity at
astrophysical distances and a comprehensive list of references we
refer to \cite{aguirre}. A possible relation to quintessence has been
speculated about in \cite{cwgal}.

\vspace*{2\baselinestretch pt}
In the present paper we are going to explore the idea that the IR
quantum effects of QEG, in particular the growth of $G$ at large
distances, induces a modified Newtonian potential similar to \eqref{2}
or \eqref{3}. If so, one can perhaps solve the missing mass problem in
a very elegant and ``minimal'' manner by simply quantizing the fields
which are known to exist anyhow, without having to introduce ``dark
matter'' on an ad hoc basis.\footnote{See refs.\ 
  \cite{bertodm,bertoproc} for a related analysis within a
  perturbatively renormalizable higher derivative gravity.}

It is particularly intriguing that the potentials we would like to
derive within QEG are strikingly similar to the nonperturbative
quark--antiquark potentials generated by (quenched) QCD. In particular
eqs.\ \eqref{1} and \eqref{3} are mathematically identical, describing
``linear confinement'', while the MOND potential increases slightly
more slowly at large distances. In view of the many similarities
between QCD and QEG it is hard to believe that this should be a mere
coincidence.

The purpose of the present paper is to learn as much as possible about
the gravitational RG trajectory Nature has chosen and to investigate
the possibility that the IR renormalization effects of QEG are ``at
work'' at galactic and cosmological scales.

The remaining sections of the paper are organized as follows. In
Section \ref{s2} we discuss the Einstein--Hilbert truncation of theory
space with an emphasis on the strong IR renormalization effects it
gives rise to. In Sections \ref{s3} and \ref{s4}, we analyze the RG
trajectory realized in Nature, first using mostly analytical results
(Section \ref{s3}) and then also phenomenological input (Section
\ref{s4}). In Section \ref{s5} we employ a plausible model of the
trajectory in the deep IR to demonstrate that an extremely tiny
variation of Newton's constant would explain the observed flat
rotation curves. The results are summarized in Section \ref{s6}.
%
%
%
%
%
%
%
%
\section{Towards the infrared with the \\Einstein--Hilbert truncation}
\label{s2}
\subsection{Structure of the RG flow} \label{s2.1}
In this subsection we discuss some properties of the Einstein--Hilbert
truncation, in particular the classification of its RG trajectories.
The emphasis will be on their behavior in the IR. We refer to
\cite{mr} and \cite{frank1} for further details.

Our basic tool is the effective average action $\Gamma_{k} \bigl[
g_{\mu \nu} \bigr]$, a free energy functional which depends on the
metric and a momentum scale $k$ with the interpretation of a variable
infrared cutoff. The action $\Gamma_{k}$ is similar to the ordinary
effective action $\Gamma$ which it approaches for $k \to 0$. The main
difference is that the path integral defining $\Gamma_{k}$ extends
only over quantum fluctuations with covariant momenta $p^{2} > k^{2}$.
The modes with $p^{2} < k^{2}$ are given a momentum dependent
$(\text{mass})^{2} \propto R_{k} \bigl( p^{2} \bigr)$ and are
suppressed therefore. As a result, $\Gamma_{k}$ describes the dynamics
of metrics averaged over spacetime volumes of linear dimension
$k^{-1}$.  The functional $\Gamma_{k} \bigl[ g_{\mu \nu} \bigr]$ gives
rise to an effective field theory valid near the scale $k$. Hence,
when evaluated at tree level, $\Gamma_{k}$ correctly describes all
quantum gravitational phenomena, including all loop effects, provided
the typical momentum scales involved are all of order $k$. 

Considered a function of $k$, $\Gamma_{k}$ describes a RG trajectory
in the space of all action functionals. The trajectory can be obtained
by solving an exact functional RG equation. In practice one has to
resort to approximations. Nonperturbative approximate solutions can be
obtained by truncating the space of action functionals, i.\,e.\ by
projecting the RG flow onto a finite--dimensional subspace which
encapsulates the essential physics.

The ``Einstein--Hilbert truncation'', for instance, approximates
$\Gamma_{k}$ by a linear combination of the monomials $\int
\!\!\sqrt{g\,} \, R$ and $\int \!\!\sqrt{g\,}$. Their prefactors
contain the running Newton constant $G (k)$ and the running
cosmological constant $\Lambda (k)$. Their $k$--dependence is governed
by a system of two coupled ordinary differential equations.

The flow equations resulting from the Einstein--Hilbert truncation are
most conveniently written down in terms of the dimensionless
``couplings'' $g(k) \equiv k^{d-2} \, G(k)$ and $\lambda(k) \equiv
\Lambda(k) / k^{2}$ where $d$ is the dimensionality of spacetime.
Parameterizing the RG trajectories by the ``RG time'' $t \equiv \ln k$
the coupled system of differential equations for $g$ and $\lambda$
reads $\partial_{t} \lambda = \boldsymbol{\beta}_{\lambda}$,
$\partial_{t} g = \boldsymbol{\beta}_{g}$, where the
$\boldsymbol{\beta}$--functions are given by
\begin{align}
\begin{split}
\boldsymbol{\beta}_{\lambda}(\lambda, g) 
& = 
-(2-\eta_{\text{N}})\, \lambda + \tfrac{1}{2}\, (4 \pi)^{1-d/2}  \, g \\
& \phantom{{==}}
\times \left[ 2 \, d(d+1) \, \Phi^1_{d/2}(-2\lambda)
- 8 \, d \, \Phi^1_{d/2}(0) 
- d(d+1) \, \eta_{\text{N}} \, \widetilde{\Phi}^1_{d/2}(-2 \lambda) \right]  
\\
\boldsymbol{\beta}_g(\lambda, g) 
& = 
\left(d-2+\eta_{\text{N}} \right) \, g
\end{split}
\label{10}
\end{align}
Here $\eta_{\text{N}}$, the anomalous dimension of the operator $\int
\!\! \sqrt{g\,} \,R$, has the representation
\begin{align}
\eta_{\text{N}}(g, \lambda) = \frac{g \, B_1(\lambda)}{1-g \, B_2(\lambda) }.
\label{11}
\end{align}
The functions $B_1(\lambda)$ and $B_2(\lambda)$ are defined by
\begin{align}
\begin{split}
B_1(\lambda) & \equiv 
\tfrac{1}{3} \,(4 \pi)^{1-d/2} 
\biggl[ d(d+1) \, \Phi^1_{d/2-1}(-2\lambda) 
- 6 \, d(d-1) \, \Phi^2_{d/2}(-2\lambda) \biggr. \\
 & \phantom{{==} \frac{1}{3} \,(4 \pi)^{1-d/2} \biggl[ \biggr.]} \biggl.
-4 \, d \,\Phi^1_{d/2-1}(0) - 24 \Phi^2_{d/2}(0) \bigg] 
\\
B_2(\lambda) &\equiv
-\tfrac{1}{6} \,(4 \pi)^{1-d/2} \, 
\left[d(d+1) \, \widetilde{\Phi}^1_{d/2-1}(-2\lambda)
-6 \, d(d-1) \, \widetilde{\Phi}^2_{d/2}(-2\lambda) \right].
\end{split}
\label{12}
\end{align}
The above expressions contain the ``threshold functions''
$\Phi^{p}_{n}$ and $\widetilde \Phi^{p}_{n}$. They are given by
\begin{align}
\Phi^p_n(w) &= 
\frac{1}{\Gamma(n)} \int \limits^{\infty}_{0} \!\! \text{d}z~
z^{n-1} \, \frac{ R^{(0)}(z) - z\, R^{(0)\prime}(z)}{\left[ z +  R^{(0)}(z) + w \right]^p\,}
\label{13}
\end{align}
and a similar formula for $\widetilde \Phi^{p}_{n}$ without the
$R^{(0)\prime}$--term. In fact, $R^{(0)}$ is a dimensionless version of
the cutoff function $R_{k}$, i.\,e.\ $R_{k} \bigl( p^{2} \bigr)
\propto k^{2} \, R^{(0)} \bigl( p^{2} / k^{2} \bigr)$. Eq.\ \eqref{13}
shows that $\Phi^{p}_{n} (w)$ becomes singular for $w \to -1$. (For
all admissible cutoffs, $z + R^{(0)} (z)$ assumes its minimum value
$1$ at $z=0$ and increases monotonically for $z > 0$.) If $\lambda
>0$, the $\Phi$'s in the $\boldsymbol{\beta}$--functions are evaluated
at negative arguments $w \equiv - 2 \lambda$. As a result, the
$\boldsymbol{\beta}$--functions diverge for $\lambda \nearrow 1/2$ and
the RG equations define a flow on a half--plane only: $- \infty < g <
\infty$, $- \infty < \lambda < 1/2$.

This point becomes particularly
clear if one uses a sharp cutoff \cite{frank1}. Then the $\Phi$'s
either display a pole at $w = -1$,
\begin{align}
\Phi^{p}_{n} (w) = 
\frac{1}{\Gamma(n)} \, \frac{1}{p-1} \, 
\frac{1}{(1+w)^{p-1}} \quad \text{for $p>1$},
\label{14}
\end{align}
or, in the special case $p=1$, they have a logarithmic singularity at $w=-1$:
\begin{align}
\Phi^{1}_{n} (w) =
- \Gamma (n)^{-1} \, \ln (1+w) + \varphi_{n}.
\label{15}
\end{align}
The constants $\varphi_{n} \equiv \Phi^{1}_{n} (0)$ parameterize the
residual cutoff scheme dependence which is still present after having
opted for a sharp cutoff. In numerical calculations we shall take them
equal to the corresponding $\Phi^{1}_{n} (0)$--value of a smooth
exponential cutoff\footnote{For this purpose we employ the exponential
  cutoff with ``shape parameter'' $s=1$. In $d=4$, the only
  $\varphi$'s we need are $\varphi_{1} = \zeta (2)$ and $\varphi_{2} =
  2 \, \zeta (3)$.  See ref.\ \cite{frank1} for a detailed discussion
  of the sharp cutoff.}, but their precise value has no influence on
the qualitative features of the RG flow \cite{frank1}. The
corresponding $\widetilde \Phi$'s are constant for the sharp cutoff:
$\widetilde \Phi^{1}_{n} (w) = \delta_{p1} / \Gamma (n+1)$.

From now on we continue the discussion in $d=4$ dimensions. Then, with
the sharp cutoff, the coupled RG equations assume the following form:
\begin{subequations} \label{16}
\begin{align}
\partial_{t} \lambda & =
- \left( 2 - \eta_{\text{N}} \right) \, \lambda
- \frac{g}{\pi} \, \left[ 5 \, \ln (1 - 2 \, \lambda) - \varphi_{2} 
+ \frac{5}{4} \, \eta_{\text{N}} \right]
\label{16a}
\\
\partial_{t} g & = \left( 2 + \eta_{\text{N}} \right) \, g
\label{16b}
\\
\eta_{\text{N}} & =
- \frac{2 \, g}{6 \pi + 5 \, g} \,
\left[ \frac{18}{1 - 2 \, \lambda} 
+ 5 \, \ln (1 - 2 \, \lambda)
- \varphi_{1} + 6 \right].
\label{16c}
\end{align}
\end{subequations}

\begin{figure}
\centering
\shadowbox{
\begin{minipage}{0.875\textwidth}
\setcaptionwidth{0.875\textwidth}
\centering
\includegraphics[width=0.95\textwidth]{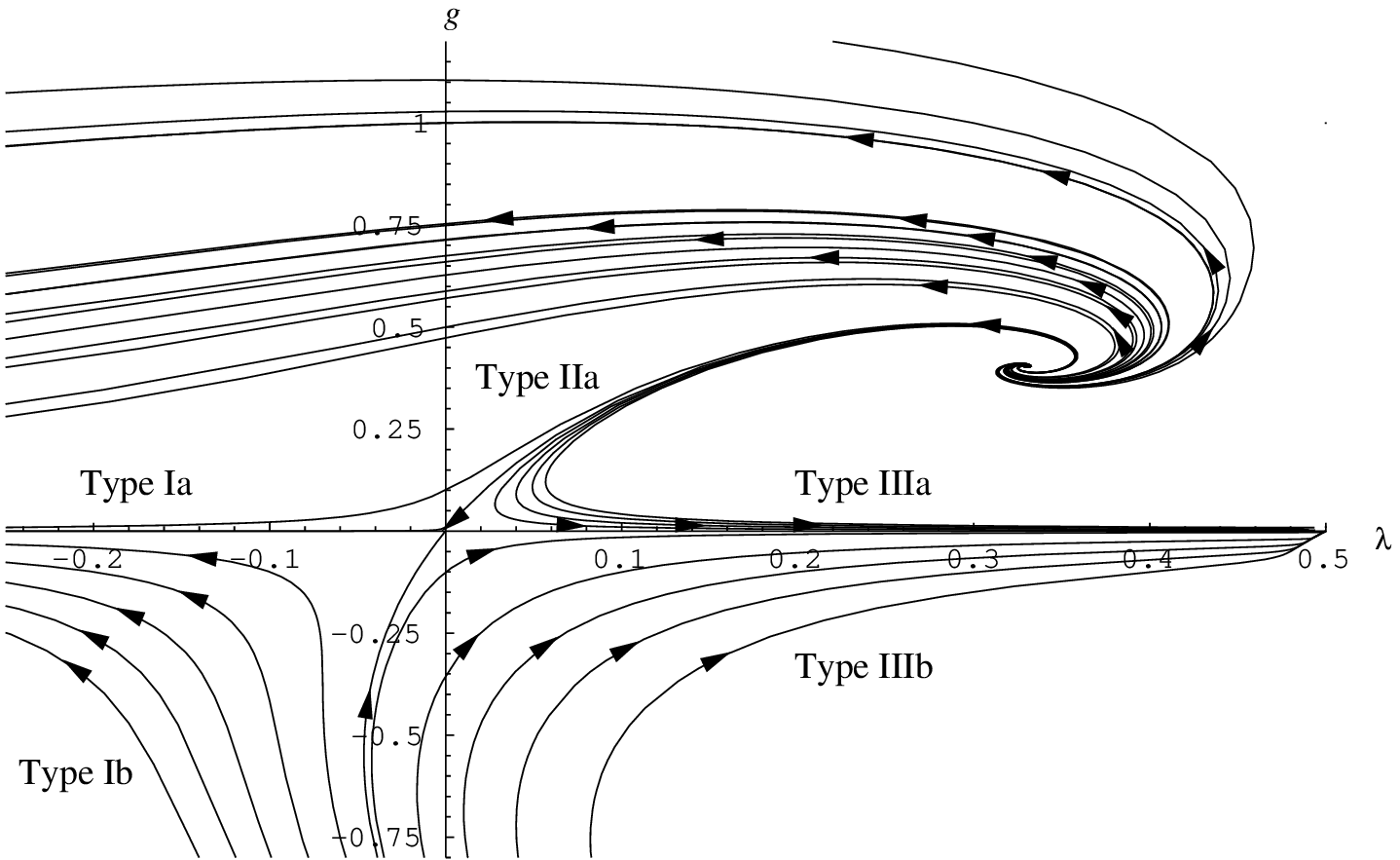}
\caption{RG flow on the $g$-$\lambda$--plane. The arrows point in the direction of decreasing values of $k$. (From ref.\ \cite{frank1}.)}
\label{fig1}
\end{minipage}}
\end{figure}

Solving the system \eqref{16} numerically \cite{frank1} we obtain the
phase portrait shown in Fig.\ \ref{fig1}.
%
%
%
%
%
The RG flow is dominated by two fixed points $\left( g_{*}, \lambda_{*}
\right)$: a Gaussian fixed point (GFP) at $g_{*} = \lambda_{*} =0$,
and a non--Gaussian fixed point (NGFP) with $g_{*} >0$ and $\lambda_{*}
>0$. There are three classes of trajectories emanating from the NGFP:
trajectories of Type Ia and IIIa run towards negative and positive
cosmological constants, respectively, and the single trajectory of
Type IIa (``separatrix'') hits the GFP for $k\to 0$. The
short--distance properties of QEG are governed by the NGFP; for $k \to
\infty$, in Fig.\ \ref{fig1} all RG trajectories on the half--plane
$g>0$ run into this point. The conjectured nonperturbative
renormalizability of QEG is due to the NGFP: if it is present in the
full RG equations, it can be used to construct a microscopic quantum
theory of gravity by taking the limit of infinite UV cutoff along one
of the trajectories running into the NGFP, thus being sure that the
theory does not develop uncontrolled singularities at high energies
\cite{wein}. By definition, QEG is the theory whose bare action $S$
equals the fixed point action $\lim_{k \to \infty} \Gamma_{k} \bigl[
g_{\mu \nu} \bigr]$.

The trajectories of Type IIIa have an important property which is not
resolved in Fig.\ \ref{fig1}. Within the Einstein--Hilbert
approximation they cannot be continued all the way down to the
infrared ($k=0$) but rather terminate at a finite scale
$k_{\text{term}} >0$. At this scale they hit the singular boundary
$\lambda = 1/2$ where the $\boldsymbol{\beta}$--functions diverge. As a result, the flow
equations cannot be integrated beyond this point. The value of
$k_{\text{term}}$ depends on the trajectory considered.

In ref.\ \cite{frank1} the behavior of $g$ and $\lambda$ close to the
boundary was studied in detail. The aspect which is most interesting
for the present discussion is the following. As the trajectory gets
close to the boundary, $\lambda$ approaches $1/2$ from below. In this
domain the anomalous dimension \eqref{16c} is dominated by its pole
term:
\begin{align}
\eta_{\text{N}} \approx - \frac{36 \, g}{6 \pi + 5 \, g} \,
\frac{1}{1 - 2 \, \lambda}.
\label{17}
\end{align}
Obviously $\eta_{\text{N}} \searrow - \infty$ for $\lambda \nearrow
1/2$, and eventually $\eta_{\text{N}} = - \infty$ at the boundary.
This behavior has a dramatic consequence for the (dimensionful) Newton
constant. Since $\partial_{t} G = \eta_{\text{N}} \, G$, the large and
negative anomalous dimension causes $G$ to grow very strongly when $k$
approaches $k_{\text{term}}$ from above. This behavior is sketched
schematically in Fig.\ \ref{fig2}.
%
%
%
%
%
%
\begin{floatingfigure}{0.62\textwidth}
\shadowbox{
\begin{minipage}{0.52\textwidth}
\setcaptionwidth{0.95\textwidth}
\includegraphics[width=0.95\textwidth]{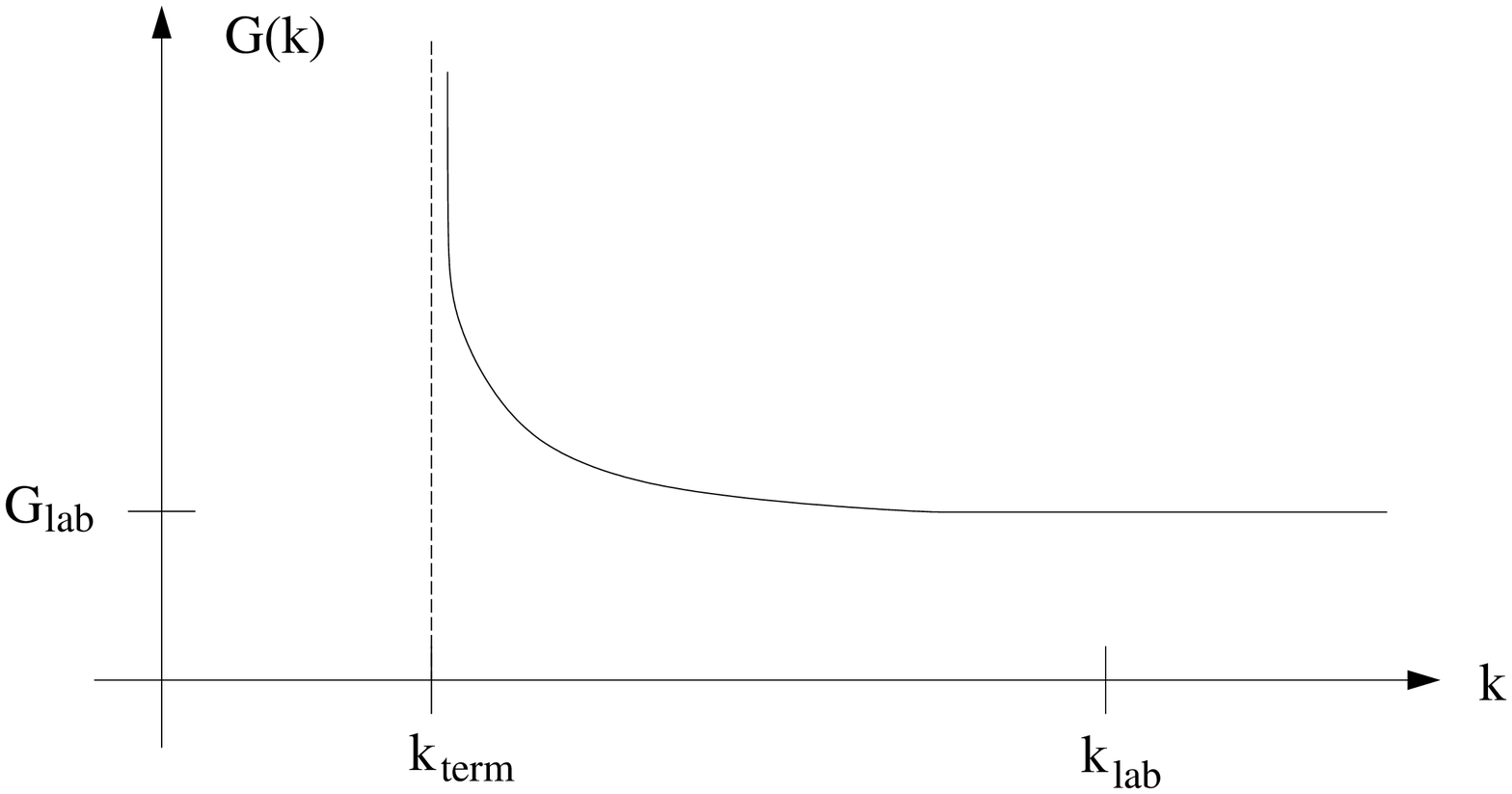}
\caption{Schematic behavior of $G (k)$ for trajectories of
Type IIIa.} 
\label{fig2}
\end{minipage}}
\end{floatingfigure}
%
%
%
%
%
At moderately large scales $k$, well below the NGFP regime, $G$ is
approximately constant. As $k$ is lowered towards $k_{\text{term}}$,
$G (k)$ starts growing because of the pole in $n_{\text{N}} \propto 1
/ \left( 1 - 2 \, \lambda \right)$, and finally, at
$k=k_{\text{term}}$, it develops a vertical tangent, $\left( \text{d}
  G / \text{d} k \right) \bigl( k_{\text{term}} \bigr) = - \infty$.
The cosmological constant is finite at the termination point: $\Lambda
\bigl( k_{\text{term}} \bigr) = k_{\text{term}}^{2} / 2$.

By fine--tuning the parameters of the trajectory the scale
$k_{\text{term}}$ can be made as small as we like.

Since it happens only ve\-ry close to $\lambda = 1/2$, the divergence at
$k_{\text{term}}$ is not visible on the scale of Fig.\ \ref{fig1}.
(Note also that $g$ and $G$ are related by a decreasing factor of
$k^{2}$.)

The phenomenon of trajectories which terminate at a finite scale is
not special to gravity, it occurs also in truncated flow equations of
theories which are understood much better. Typically it is a symptom
which indicates that the truncation used becomes insufficient at small
$k$. In QCD, for instance, thanks to asymptotic freedom, simple local
truncations are sufficient in the UV, but a reliable description in
the IR requires many complicated (nonlocal) terms in the truncation
ansatz. Thus the conclusion is that for trajectories of Type IIIa the
Einstein--Hilbert truncation is reliable only well above
$k_{\text{term}}$. It is to be expected, though, that in an improved
truncation those trajectories can be continued to $k=0$.

We believe that while the Type IIIa trajectories of the
Einstein--Hilbert truncation become unreliable very close to
$k_{\text{term}}$, their prediction of a growing $G (k)$ for
decreasing $k$ in the IR is actually correct. The function $G (k)$
obtained from the differential equations \eqref{16} should be
reliable, at least at a qualitative level, as long as $\lambda \ll 1$.
For special trajectories the IR growth of $G (k)$ sets in at extremely
small scales $k$ only. Later on we shall argue on the basis of
a gravitational ``RG improvement''
\cite{bh,cosmo1,cosmo2,elo,esposito,scalfact} that this IR growth
is responsible for the non--Keplerian rotation curves observed in
spiral galaxies.

The other trajectories with $g>0$, the Types Ia and IIa, do not
terminate at a finite scale. The analysis of ref.\ \cite{frank1}
suggests that they are reliably described by the Einstein--Hilbert
truncation all the way down to $k=0$.
\subsection{What drives the IR renormalization?} \label{s2.2}
Next we discuss a simple physical argument which sheds light on the
dynamical origin of the expected strong IR effects. As we shall see,
they are due to an ``instability driven renormalization'', a
phenomenon well known from many other physical systems
\cite{oliver0,polinst}, spontaneous symmetry breaking being the prime
example.

For an arbitrary set of fields $\phi$, and in a slightly symbolic
notation\footnote{We ignore possible complications due to gauge
  invariance. They are inessential for the present discussion.}, the
exact RG equation for the effective average action reads
\cite{avact,avactrev}
\begin{align}
\partial_{t} \Gamma_{k} [\phi] = \tfrac{1}{2} \, 
\text{Tr} \left[ 
\bigl( \Gamma^{(2)}_{k} [\phi] + R_{k} \bigr)^{-1} \, \partial_{t} R_{k} 
\right].
\label{18}
\end{align}
It contains the fully dressed effective propagator $\bigl(
\Gamma^{(2)}_{k} + R_{k} \bigr)^{-1}$ where $\Gamma^{(2)}_{k}$ denotes
the Hessian of $\Gamma_{k}$ and $R_{k}$ the cutoff operator. It is
instructive to rewrite \eqref{18} in the form
\begin{align}
\partial_{t} \Gamma_{k} [\phi] = \frac{1}{2} \, \frac{D}{D t} \, 
\ln \det \bigl( \Gamma^{(2)}_{k} [\phi] + R_{k} \bigr)
\label{12-1}
\end{align}
where the derivative $D / D t$ acts on the $k$--dependence of $R_{k}$
only. The RHS of \eqref{12-1} represents a
``$\boldsymbol{\beta}$--functional'' which summarizes all the
infinitely many ordinary $\boldsymbol{\beta}$--functions in a compact
way. Obviously its essential ingredient is a kind of a ``one--loop
determinant''. It differs from that of a standard one--loop
calculation by the presence of the cutoff term $R_{k}$ and by the use
of the dressed inverse propagator $\Gamma^{(2)}_{k}$ rather than the
classical $S^{(2)}$. More importantly, in the standard situation one
expands the classical action $S [\phi]$ about its minimum
$\phi_{\text{min}} (x)$, in which case the one--loop effective action
$\ln \det \bigl( S^{(2)} [\phi_{\text{min}}] \bigr)$ sums up the
zero--point energies of the small stable oscillations about
$\phi_{\text{min}} (x)$. On the RHS of \eqref{12-1}, instead, $\phi
(x)$ is a prescribed external field, the argument of $\Gamma_{k}$ on
the LHS. It can be changed freely, eq.\ \eqref{12-1} holds for
\textit{all} $\phi (x)$, so that $\phi (x)$ is not in general a
stationary point of $\Gamma_{k}$.

Thus we may conclude that the basic physical mechanism which drives
the RG flow is that of quantum fluctuations on arbitrary
\textit{off--shell} backgrounds $\phi$. They determine the
$\boldsymbol{\beta}$--functional, and depending on how ``violent''
those fluctuations are, the RG running is weaker or stronger.

It is helpful to consider two extreme cases. Let us first assume that,
for a certain fixed $\phi (x)$, the operator $\Gamma^{(2)}_{k}$ is
positive definite; then $\Gamma^{(2)}_{k} + R_{k}$ is positive,
too,\footnote{In order to capture the essence of the argument it is
  sufficient to use a mass--type cutoff \cite{avactrev} for which
  $R_{k} = k^{2}$ and $R^{0} \left( p^{2} / k^{2} \right) =1$ for all
  $p^{2}$.} and the quadratic action governing the fluctuations
$\delta \phi (x)$ about $\phi (x)$ given by $\int \!\! \text{d}^{4}
x~\delta \phi \, \bigl( \Gamma^{(2)}_{k} + k^{2} \bigr) \, \delta
\phi$ is a positive quadratic form. In this case the computation of
the $\boldsymbol{\beta}$--functional amounts to summing up the
zero--point energies of small stable fluctuations which would not grow
unboundedly. The ``fluctuation induced'' renormalizations of the
parameters in $\Gamma_{k}$ which they give rise to are comparatively
weak.

If, on the other hand, $\Gamma^{(2)}_{k} [\phi]$ has one or several
negative eigenvalues $\mu < 0$ with $|\mu| < \infty$, then
$\Gamma^{(2)}_{k} + R_{k} \equiv \Gamma^{(2)}_{k} + k^{2}$ is positive
only in presence of the IR regulator, for $k^{2} > |\mu|$. Without the
IR regulator there is a real physical instability. Within the linear
approximation the fluctuation modes grow unboundedly; beyond the
linear approximation they would try to ``condense'' in order to turn
$\phi$ into a stable ground state. Following a well behaved RG
trajectory one stays in the regime $k^{2} > |\mu|$ where
$\Gamma^{(2)}_{k} + k^{2}$ is positive. However, when $k^{2}$
approaches $|\mu|$ from above, the lowest eigenvalue of
$\Gamma^{(2)}_{k} + k^{2}$ gets very close to zero, and the RG flow is
strongly affected by the presence of the nearby singularity. The
effective propagator $\bigl( \Gamma^{(2)}_{k} + R_{k} \bigr)^{-1}$
becomes very large, and typically this leads to an enormous growth of
the (standard) $\boldsymbol{\beta}$--functions when $k^{2} \searrow
|\mu|$. They give rise to comparatively strong ``instability induced''
renormalizations. We shall see in a moment that the IR effects of QEG
are precisely of this type.

In order to find the RG flow on the full theory space the above
stability analysis and the ``summation of zero--point energies'' has
to be performed for infinitely many different backgrounds $\phi (x)$;
they are needed in order to ``project out'' all the possible field
monomials which constitute the functional $\Gamma_{k}$. On a truncated
theory space just a few $\phi$'s might be sufficient.

In order to illustrate the relationship between the (ordinary)
$\boldsymbol{\beta}$--functions and the instability presented by
$\Gamma^{(2)}_{k}$ let us look at a scalar model (on flat spacetime)
in a simple truncation:
\begin{align}
\Gamma_{k} [\phi] = \int \!\! \text{d}^{4} x~
\Big \{
\tfrac{1}{2} \, \partial_{\mu} \phi \, \partial^{\mu} \phi
+ \tfrac{1}{2} \, m^{2} (k) \, \phi^{2}
+ \tfrac{1}{12} \, \lambda (k) \, \phi^{4}
\Big \}.
\label{12-2}
\end{align}
Here $\phi$ denotes a real, $\mathcal{Z}_{2}$--symmetric scalar field, and the
truncation ansatz \eqref{12-2} retains only a running mass and
$\phi^{4}$--coupling. In a momentum basis where $- \partial_{\mu}
\partial^{\mu} \overset{\wedge}{=} p^{2} > 0$ we have
\begin{align}
\Gamma^{(2)}_{k} \overset{\wedge}{=} p^{2} + m^{2} (k) + \lambda (k) \, \phi^{2}.
\label{12-3}
\end{align}
Always assuming that $\lambda >0$, we see that $\Gamma^{(2)}_{k}$ is
positive if $m^{2} >0$; but when $m^{2} <0$ it can become negative for
$\phi^{2}$ small enough. Of course, the negative eigenvalue for $\phi
=0$, for example, indicates that the fluctuations want to grow, to
``condense'', and thus to shift the field from the ``false vacuum'' to
the true one. By the mechanism discussed above, this gives rise to
strong instability induced renormalizations. In fact, the standard
$\boldsymbol{\beta}$--functions for $m^{2}$ and $\lambda$ can be found
by inserting \eqref{12-3} into \eqref{18}, taking two and four
derivatives with respect to $\phi$, respectively, and then setting
$\phi=0$ in order to project out $\partial_{t} m^{2}$ and
$\partial_{t} \lambda$. As a result, the
$\boldsymbol{\beta}$--functions are given by $p$--integrals over
(powers of) the propagator
\begin{align}
\Bigl[ p^{2} + m^{2} (k) + k^{2} \Bigr]^{-1}.
\label{12-4}
\end{align}
In the symmetric phase ($m^{2} >0$) this (euclidean!) propagator has
no pole, and the resulting $\boldsymbol{\beta}$--functions are
relatively small. In the broken phase ($m^{2} <0$), however, there is
a pole at $p^{2} = - m (k)^{2} - k^{2}$ provided $k^{2}$ is small
enough: $k^{2} < | m (k)^{2}|$. For $k^{2} \searrow | m (k)^{2}|$ the
$\boldsymbol{\beta}$--functions become large and there are strong
instability induced renormalizations.

In a reliable truncation, a physically realistic RG trajectory in the
spontaneously broken regime will not hit the singularity at $k^{2} = |
m (k)^{2}|$, but rather make $m (k)$ run in precisely such a way that
$| m (k)^{2}|$ is always smaller than $k^{2}$. This requires that
\begin{align}
- m (k)^{2} \propto k^{2}.
\label{12-5}
\end{align}
This strong instability induced mass renormalization is necessary in
order to evolve an originally $W$--shaped symmetry breaking classical
potential into an effective potential which is convex and has a flat
bottom. (See \cite{avactrev} for a detailed discussion of this point.)

Unfortunately the two--parameter truncation \eqref{12-2} is too
rudimentary for a reliable description of the broken phase. Its RG
trajectories actually do run into the singularity. They terminate at a
finite scale $k_{\text{term}}$ with $k_{\text{term}}^{2} = | m
(k_{\text{term}})^{2}|$ at which the $\boldsymbol{\beta}$--functions
diverge. Instead, if one allows for an arbitrary running potential
$U_{k} (\phi)$, containing infinitely many couplings, all trajectories
can be continued to $k=0$, and for $k \searrow 0$ one finds indeed the
quadratic mass renormalization \eqref{12-5} \cite{avactrev}.

Let us return to gravity now where $\phi$ corresponds to the metric.
In the Einstein--Hilbert truncation it suffices to insert the metric
corresponding to a sphere $\mathrm{S}^{4} (r)$ of arbitrary radius
$r$ into the flow equation\footnote{We stress, however, that the
  $\boldsymbol{\beta}$--functions do not depend on the choice of
  background used for projecting onto the various invariants
  \cite{avactrev,mr,oliver1}.} in order to disentangle the
contributions from the two invariants $\int \! \text{d}^{4} x 
\sqrt{g\,} \propto r^{4}$ and $\int \! \text{d}^{4} x \sqrt{g\,}
\, R \propto r^{2}$. Thus we may think of the Einstein--Hilbert flow
as being a manifestation of the dynamics of graviton fluctuations on
$\mathrm{S}^{4} (r)$. This family of backgrounds, labeled by $r$, is
``off--shell'' in the sense that $r$ is completely arbitrary and not
fixed by Einstein's equation in terms of $\Lambda$.

It is convenient to decompose the fluctuation $h_{\mu \nu}$ on the
sphere into irreducible (TT, TL, $\cdots$) components \cite{oliver1}
and to expand the irreducible pieces in terms of the corresponding
spherical harmonics. For $h_{\mu \nu}$ in the transverse--traceless
(TT) sector, say, the operator $\Gamma^{(2)}_{k} + R_{k}$ equals, up
to a positive constant,
\begin{align}
- D^{2} +  8 \, r^{-2} + k^{2} - 2 \, \Lambda (k)
\label{12-6}
\end{align}
with $D^{2} \equiv g^{\mu \nu} \, D_{\mu} D_{\nu}$ the covariant
Laplacian acting on TT tensors. The spectrum of $- D^{2}$, denoted $\{
p^{2} \}$, is discrete and positive. Obviously \eqref{12-6} is a
positive operator if the cosmological constant is negative. In this
case there are only stable, bounded oscillations, leading to a mild
fluctuation induced renormalization. This is precisely what we observe
in the IR of the Type Ia trajectories: there is virtually no
non--canonical parameter running below $k = m_{\text{Pl}}$. The
situation is very different for $\Lambda > 0$ where, for $k^{2}$
sufficiently small, \eqref{12-6} has negative eigenvalues, i.\,e.\
unstable eigenmodes. In fact, expanding the RHS of the flow equation
to orders $r^{2}$ and $r^{4}$ the resulting
$\boldsymbol{\beta}$--functions are given by traces (spectral sums)
containing the propagator \cite{mr}
\begin{align}
\left[ p^{2} + k^{2} - 2 \, \Lambda (k) \right]^{-1}.
\label{19-1}
\end{align}
The crucial point is that the propagator \eqref{19-1} can have a pole
when $\Lambda (k)$ is too large and positive. It occurs for $\Lambda
(k) \geq k^{2}/2$, or equivalently $\lambda (k) \geq 1/2$, at $p^{2} =
2 \, \Lambda (k) - k^{2}$. Upon performing the $p^{2}$--sum this pole
is seen to be responsible for the terms $\propto 1 / \left( 1 - 2 \,
  \lambda \right)$ and $\ln (1 - 2 \, \lambda)$ in the
$\boldsymbol{\beta}$--functions which become singular at $\lambda =
1/2$. The allowed part of the $g$-$\lambda$--plane ($\lambda < 1/2$)
shown in Fig.\ \ref{fig1} corresponds to the situation $k^{2} > 2 \,
\Lambda (k)$ where the singularity is avoided thanks to the large
regulator mass. When $k^{2}$ approaches $2 \, \Lambda (k)$ from above
the $\boldsymbol{\beta}$--functions become large and strong
renormalizations set in, driven by the modes which would go
unstable\footnote{From the propagator \eqref{19-1} it is obvious that
  the smaller the eigenvalue $p^{2}$ of a fluctuation mode, the higher
  is the scale $k^{2}$ at which this particular mode starts
  contributing significantly, and the more important is its impact on
  the RG flow. A galaxy--size fluctuation is more important than a
  solar system--size fluctuation, for example.} at $k^{2} = 2 \,
\Lambda$.

In this respect the situation is completely analogous to the scalar
theory discussed above: Its symmetric phase ($m^{2} >0$) corresponds
to gravity with $\Lambda <0$; in this case all fluctuation modes are
stable and only small renormalization effects occur. Conversely, in
the broken phase ($m^{2} <0$) and in gravity with $\Lambda >0$, there
are modes which are unstable in absence of the IR regulator. They lead
to strong IR renormalization effects for $k^{2} \searrow | m (k)^{2}|$
and $k^{2} \searrow 2 \, \Lambda (k)$, respectively. The gravitational
Type Ia (Type IIIa) trajectories are analogous to those of the
symmetric (broken) phase of the scalar model.

As for the behavior of the RG trajectories near the boundary the
crucial question is whether, when $k$ is lowered, $\Lambda (k)$
decreases at least as fast as $k^{2}$ or more slowly. In the first
case the trajectory would never reach the singularity, while it does
so in the second. In the Einstein--Hilbert approximation the
trajectories of Type IIIa indeed belong to the second case; since
$\Lambda (k)$ does not decrease fast enough the RG trajectory runs
into the pole at a certain $k_{\text{term}}$ where
$k_{\text{term}}^{2} = 2 \, \Lambda (k_{\text{term}})$.

The termination of certain trajectories is not specific to
gravity. We saw that it happens also in the scalar model if we use the
over--simplified truncation \eqref{12-2}. This simple ansatz has
similar limitations as the Einstein--Hilbert truncation. In the scalar
case the cure to the problem of terminating trajectories is known
\cite{avactrev}: If one uses a more general truncation, allowing for a
non--polynomial $U_{k} (\phi)$, the RG trajectories never reach the
singularity and extend to $k=0$, with strong renormalizations,
however, in particular the quadratic running \eqref{12-5}. In a
certain sense the Einstein--Hilbert truncation has a similar status as
a \textit{polynomial} truncation for $U_{k} (\phi)$: it is not general
enough to be reliable down to $k=0$ for a positive cosmological
constant or in the broken phase, respectively. While there are
computationally manageable truncations of sufficient generality in the
scalar case it is not known which truncations would allow for a
reliable continuation of the Type IIIa trajectories below
$k_{\text{term}}$. They are likely to contain nonlocal invariants
\cite{oliver0,frank2,cwnonloc} which are hard to handle analytically.

In view of the scalar analogy it is a plausible and very intriguing
speculation that, for $k \to 0$, an improved gravitational truncation
has a similar impact on the RG flow as it has in the scalar case.
There the most important renormalization effect is the running of the
mass: $- m (k)^{2} \propto k^{2}$. If gravity avoids the singularity
in an analogous fashion the cosmological constant would run
proportional to $k^{2}$,
\begin{align}
\Lambda (k) = \lambda^{\text{IR}}_{*} \, k^{2}
\label{19-2}
\end{align}
with a constant $\lambda^{\text{IR}}_{*} < 1/2$. In dimensionless
units \eqref{19-2} reads $\lambda (k) = \lambda^{\text{IR}}_{*}$,
i.\,e.\ $\lambda^{\text{IR}}_{*}$ is a fixed point of the
$\lambda$--evolution. If the behavior \eqref{19-2} is actually
realized, the renormalized cosmological constant observed at very
large distances, $\Lambda (k \to 0)$, vanishes regardless of its bare
value. Clearly this would have an important impact on the cosmological
constant problem \cite{coscon}.
%
%
%
%
%
%
%
%
\section{The RG trajectory realized in Nature: \\the Einstein--Hilbert domain}
\label{s3}
How can we find out which one of the RG trajectories shown in Fig.\ 
\ref{fig1} is realized in Nature? As in every quantum field theory,
one has to experimentally determine the value of appropriate
``renormalized'' quantities.\footnote{Above we considered pure
  gravity, while the experimental data include renormalization effects
  due to matter fields. Since in this paper we are interested in order
  of magnitude estimates only, and we anyhow do not know the exact
  matter field content of Nature, we assume that the inclusion of
  matter does not change the general qualitative features of pure
  gravity. As for the nonperturbative renormalizability it is known
  that there exist matter systems with this property and that they are
  ``generic'' in a sense \cite{percacciperini}.} In Quantum
Electrodynamics (QED), for instance, one measures the electron's
charge and mass in a large--distance experiment, thus fixing $e (k)$
and $m_{\mathrm{e}} (k)$ at $k=0$. In QCD the point $k=0$ is
inaccessible, both theoretically and experimentally, so one uses a
``renormalization point'' at a higher scale $k>0$. In QEG the
situation is similar. In the extreme infrared ($k \to 0$) we have
neither theoretically reliable predictions nor precise experimental
determinations of the gravitational couplings $g$ and $\lambda$.
\subsection{Exploiting experimental information}
\label{s3.1}
We know that all gravitational phenomena at distance scales ranging
from terrestrial experiments to solar system measurements are well
described by standard General Relativity (GR). Since this theory is
based upon the Einstein--Hilbert action with constant values of $G$
and $\Lambda$, we can conclude that the RG evolution of those
parameters for $k$ between the related typical mass scales $\left(
  \unit[1]{meter} \right)^{-1}$ and $\left(
  \unit[1]{astronomical\text{ }unit} \right)^{-1}$, say, is negligibly
small. In this ``GR regime'' the renormalization group flow is
essentially the canonical one, i.\,e.\ $g (k) \propto k^{2}$ and
$\lambda (k) \propto 1/k^{2}$ which follows from $g (k) \equiv k^{2}
\, G (k)$, $\lambda (k) \equiv \Lambda (k) / k^{2}$ when $G,\Lambda =
const$. The corresponding RG trajectories are the hyperbolas $g
\propto 1 / \lambda$ depicted in Fig.\ \ref{fig3}. During the
$k$--interval defining the GR regime the true RG trajectory realized
in Nature must be very close to one of those hyperbolas.
%
%
%
%
%
%
\begin{floatingfigure}{0.56\textwidth}
\shadowbox{
\begin{minipage}{0.45\textwidth}
\setcaptionwidth{0.95\textwidth}
\centering
\includegraphics[width=0.95\textwidth]{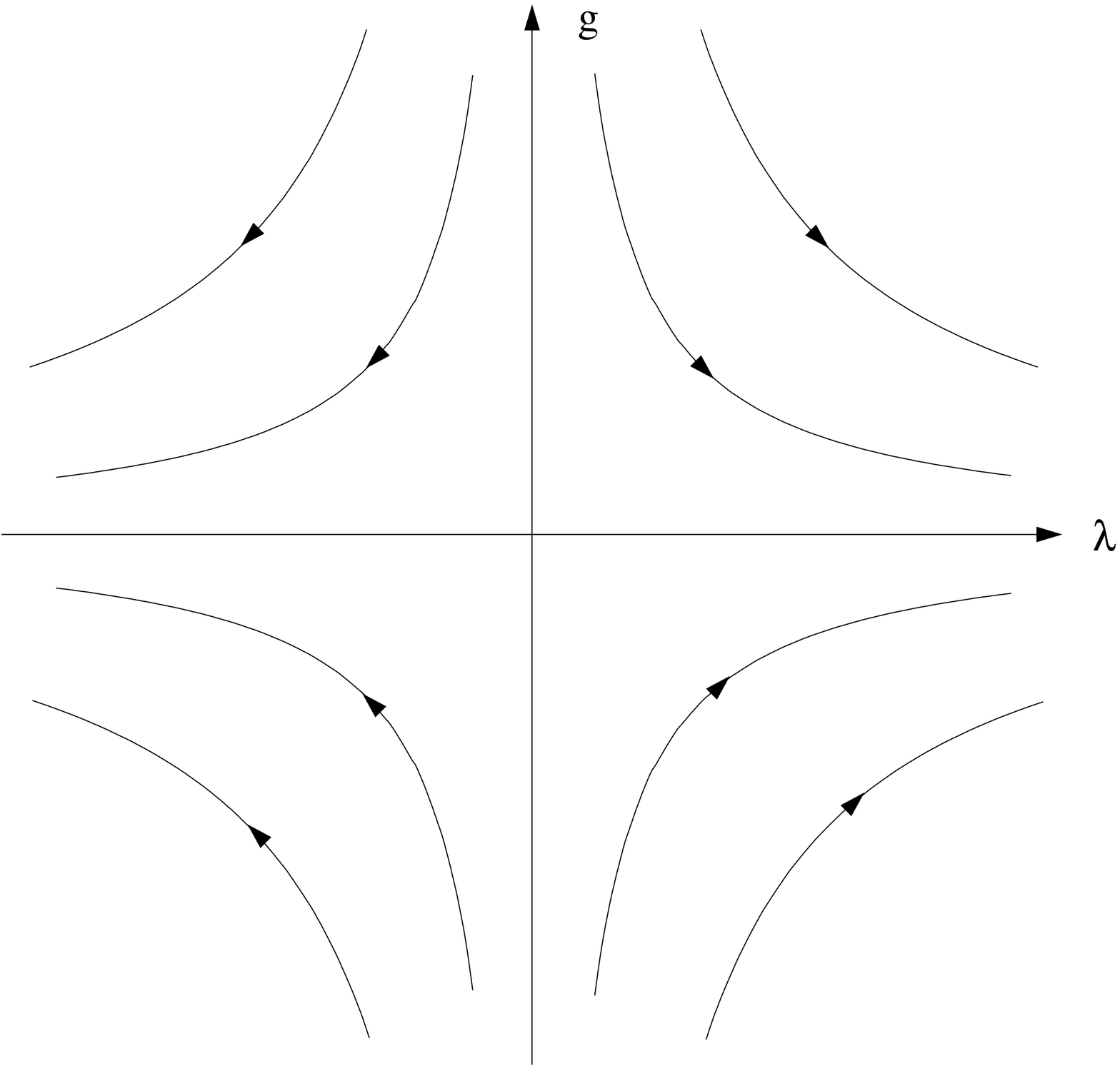}
\caption{The canonical RG flow corresponding to constant values of
the dimensionful parameters $G$ and $\Lambda$.}
\label{fig3}
\end{minipage}}
\end{floatingfigure}

Which class does this trajectory belong to? Recent CMBR and high
redshift supernova data show that the present Universe is in a state
of accelerated expansion which, in a Fried\-mann--Robertson--Walker
framework, can be explained by an nonzero \textit{positive}
cosmological constant. In the RG context this should mean that
$\Lambda \bigl( k \approx H_{0} \bigr) > 0$ since the relevant scale
is set by $H_{0}$, the present Hubble parameter. Among the
trajectories of the Einstein--Hilbert truncation only those of Type
IIIa and IIIb run towards positive $\Lambda$'s for $k \searrow 0$.
Trajectories of Type IIIb correspond to a negative $G$ and
are excluded therefore. Thus \textit{the RG trajectory realized in
  Nature, as long as it remains in the domain of validity of the
  Einstein--Hilbert truncation, belongs to Type \textnormal{IIIa}.}

We saw that trajectories of Type IIIa cannot be continued below a
certain $k_{\text{term}}$, and in the present paper we are going to
argue that $k_{\text{term}}$ is roughly of the order of typical
galaxy scales. As a result, the Einstein--Hilbert truncation is
probably insufficient to describe the (continuation of the) Type IIIa
trajectory realized in Nature at the cosmological scale $k \approx
H_{0}$ where $\Lambda$ was actually measured.  Nevertheless it seems
to be clear that, for $k$ large enough, the true RG trajectory
selected by Nature is a Einstein--Hilbert trajectory of Type IIIa. The
reason is that the other alternatives, Type Ia and Type IIa, can be
computed reliably down to $k=0$, and they do \textit{not} give rise to
a positive cosmological constant in the infrared.

\begin{figure}
\shadowbox{
\begin{minipage}{0.95\textwidth}
\setcaptionwidth{0.95\textwidth}
\centering
\includegraphics[width=0.95\textwidth]{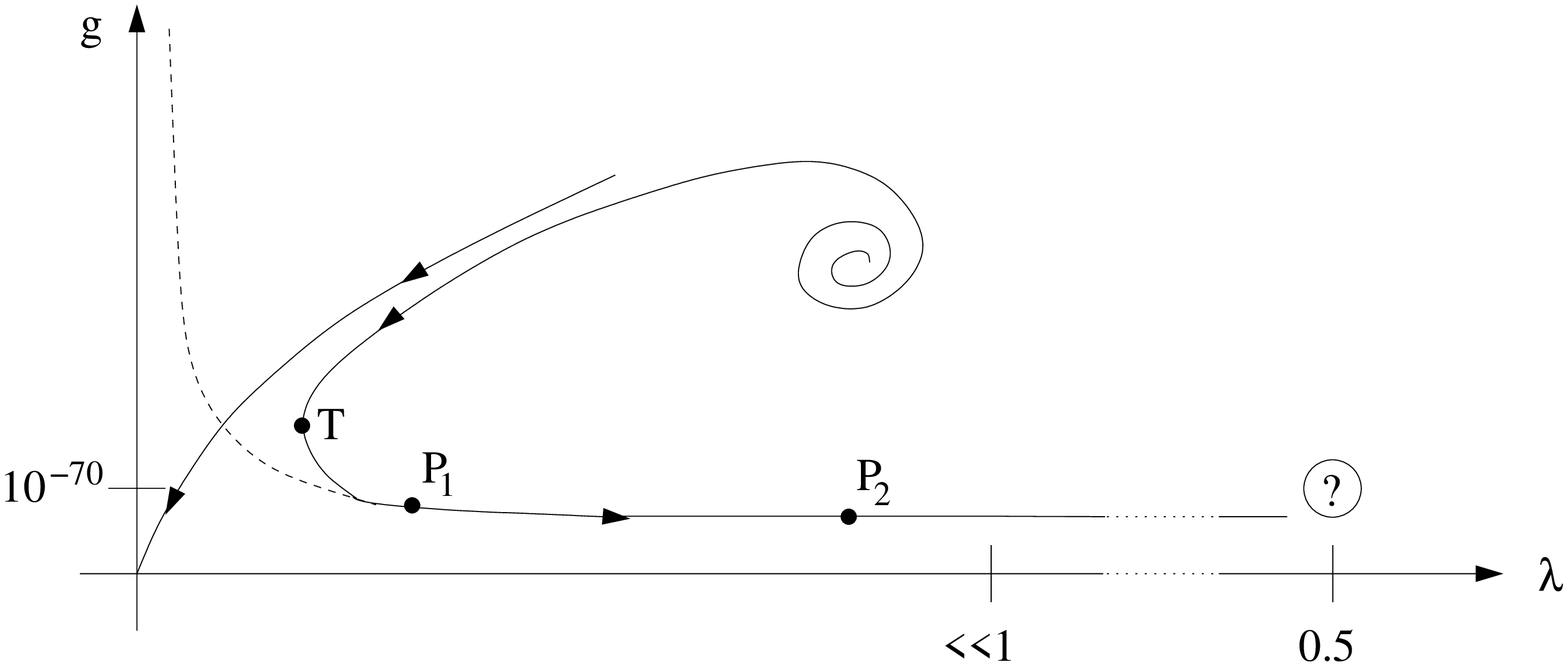}
\caption{The Type IIIa trajectory realized in Nature and the separatrix.
The dashed line is a trajectory of the canonical RG flow.} 
\label{fig4}
\end{minipage}}
\end{figure}

For our picture to be correct, the prospective Type IIIa trajectory
must contain a sufficiently long ``GR regime'' where it runs on top of
one of the hyperbolas of Fig.\ \ref{fig3}. The situation is sketched
qualitatively in Fig.\ \ref{fig4}.
%
%
%
%
%
The Type IIIa trajectory spirals out of the NGFP, approaches the
separatrix, runs almost parallel to it for a while, then ``turns
left'' near the GFP, and finally runs towards the singularity at
$\lambda = 1/2$. After the turning point where $\partial_{t} \lambda
=0$, but before it gets too close to $\lambda = 1/2$, this trajectory
is an almost perfect hyperbola of the canonical RG flow. In Fig.\ 
\ref{fig4} we indicate the latter by the dashed line which, between
the points $P_{1}$ and $P_{2}$, is indistinguishable from the Type
IIIa trajectory. It is this segment between $P_{1}$ and $P_{2}$ which
can be identified with the realm of classical GR. As we shall see in a
moment, the variation of $G$ and $\Lambda$ along the true Type IIIa
trajectory is unmeasurably small between $P_{1}$ and $P_{2}$.

Which one of the infinitely many Type IIIa trajectories did Nature
pick? We can answer this question if, from experiments or
astrophysical observations, we know a single point $\left( g, \lambda
\right)$ the trajectory passes through. Let us assume we measure the
(dimensionful) Newton constant and cosmological constant in a
``laboratory'' with a typical linear dimension of the order $1 /
k_{\text{lab}}$. We interpret the result of the measurements as the
running couplings evaluated at this scale: $G \bigl( k_{\text{lab}}
\bigr)$, $\Lambda \bigl( k_{\text{lab}} \bigr)$. Knowing those two
values, as well as the pertinent ``laboratory'' scale
$k_{\text{lab}}$, we can compute the dimensionless couplings:
\begin{align}
g \bigl( k_{\text{lab}} \bigr) = 
k_{\text{lab}}^{2} \, G \bigl( k_{\text{lab}} \bigr),
\quad
\lambda \bigl( k_{\text{lab}} \bigr) =
\Lambda \bigl( k_{\text{lab}} \bigr) /k_{\text{lab}}^{2} .
\label{20}
\end{align}
The pair $\bigl( g (k_{\text{lab}}), \lambda ( k_{\text{lab}}) \bigr)$
uniquely fixes a trajectory in the Einstein--Hilbert approximation. If
one uses a more general truncation, further parameters need to be
measured, of course. The first one of the eqs.\ \eqref{20} can be
rewritten in the following suggestive form:
\begin{align}
  g \bigl( k_{\text{lab}} \bigr) = \bigl( k_{\text{lab}} /
  m_{\text{Pl}} \bigr)^{2} \equiv \bigl( \ell_{\text{Pl}} /
  k_{\text{lab}}^{-1} \bigr)^{2}.
\label{21}
\end{align}
Here we defined the Planck length and mass in the usual way in terms
of the measured Newton constant $G \bigl( k_{\text{lab}} \bigr)$
according to $\ell_{\text{Pl}} \equiv m_{\text{Pl}}^{-1} = \sqrt{G
  \bigl( k_{\text{lab}} \bigr)\,}$.

Newton's constant has been measured at length scales
$k_{\text{lab}}^{-1}$ ranging from the size of terrestrial experiments
to solar system dimensions. Within the errors the result has always
been the same: $G \bigl( k_{\text{lab}} \bigr) = G_{\text{lab}} \equiv
\unit[6.67 \cdot 10^{-11}]{m^{3} \, kg^{-1} \, s^{-2}}$. We can now
calculate $g \bigl( k_{\text{lab}} \bigr)$ according to \eqref{20}. At
the typical scale of a terrestrial laboratory one finds
\begin{align}
g \bigl( k_{\text{lab}} \bigr) \approx 10^{-70}
\quad \text{for }  k_{\text{lab}}^{-1} = \unit[1]{m}
\label{22}
\end{align}
while at the solar system scale of $\unit[1]{astronomical \text{ } unit}$,
\begin{align}
g \bigl( k_{\text{lab}} \bigr) \approx 10^{-92}
\quad \text{for }  k_{\text{lab}}^{-1} = \unit[1]{AU}.
\label{23}
\end{align}
In any case $g \bigl( k_{\text{lab}} \bigr)$ is an extremely small
number for any $k_{\text{lab}}$ in the GR regime, $g \bigl(
k_{\text{lab}} \bigr) \lll 1$. Its precise value will not matter in
the following; for the sake of clarity we shall use the example $
k_{\text{lab}}^{-1} = \unit[1]{meter}$ and $g \bigl( k_{\text{lab}}
\bigr) \approx 10^{-70}$ for numerical illustration. Throughout the
discussion the length scale $k_{\text{lab}}^{-1}$ is assumed to lie in
the GR regime, ranging from terrestrial to solar system distances.

The determination of the associated $\lambda \bigl( k_{\text{lab}}
\bigr)$ is difficult; in fact, rather than at ``laboratory'' scales
$\unit[1]{m} \cdots \unit[1]{AU}$, $\Lambda$ was actually measured at
cosmological distance scales. For a first qualitative discussion the
following estimate is sufficient, however. According to the effective
vacuum Einstein equation at the scale $k_{\text{lab}}$, a cosmological
constant of magnitude $\Lambda \bigl( k_{\text{lab}} \bigr)$ leads to
the spacetime whose radius of curvature is of the order
\begin{align*}
r_{c} \approx \Lambda \bigl( k_{\text{lab}} \bigr)^{-1/2}
= \lambda \bigl( k_{\text{lab}} \bigr)^{-1/2} \, k_{\text{lab}}^{-1}.
\end{align*}
We know that in absence of matter, at $k_{\text{lab}}^{-1} =
\unit[1]{m}$, say, spacetime (when observed with a ``microscope'' of
resolution $k_{\text{lab}}^{-1}$) is flat with a very high precision,
i.\,e.\ that $r_{c}$ is much larger than the size of the
``laboratory'': $r_{c} \gg k_{\text{lab}}^{-1}$. As a consequence,
$\lambda \bigl( k_{\text{lab}} \bigr)$ must be very small compared to
unity:
\begin{align}
\lambda \bigl( k_{\text{lab}} \bigr) \ll 1.
\label{24}
\end{align}
This means in particular that, at $k_{\text{lab}}$ and in the entire GR
regime, the trajectory realized in Nature is still very far away from
the dangerous singularity at $\lambda = 1/2$ where the
Einstein--Hilbert approximation breaks down.
\subsection{Approximate RG flow near the GFP}
\label{s3.2}
Since $g, \lambda \ll 1$ in the GR regime we may neglect higher order
terms $g^{2}, g \lambda, \cdots$ in the eqs.\ \eqref{16} and obtain
the following flow equation linearized about the GFP:
\begin{subequations} \label{25}
\begin{align}
\partial_{t} \lambda & = - 2 \, \lambda + \varphi_{2} \, g / \pi
\label{25a}
\\
\partial_{t} g & = 2 \, g.
\label{25b}
\end{align}
\end{subequations}
In the ``linear regime'' where \eqref{25} is valid, only the
cosmological constant shows a non--canonical running, while the
dimensionful Newton constant does not evolve in the approximation
\eqref{25b}.

In the next to leading order $G$ runs according to $\partial_{t} G =
\eta_{\text{N}} \, G$ with $\eta_{\text{N}} = - b \, g$ proportional
to $g$. In any cutoff scheme $b$ is a positive constant of order
unity, $b = \left( 24 - \varphi_{1} \right) / 3 \pi$ for the sharp
cutoff. Hence
\begin{align}
- \eta_{\text{N}} \Big \rvert_{\text{RG regime}} \approx
10^{-70} \cdots 10^{-92}.
\label{26}
\end{align}
The smallness of these numbers explains the success of standard
General Relativity based upon the approximation $\eta_{\text{N}} =0$,
and it confirms our interpretation of the segment between the points
$P_{1}$ and $P_{2}$ in Fig.\ \ref{fig4} as the realm of classical
gravity.

To the right of the point $P_{2}$ in Fig.\ \ref{fig4}, at scales $k$
lower than those of the GR regime, the growth of $G (k)$ due to the
infrared instability sets in.

Let us return to the linearized flow equations \eqref{25}. They are
applicable whenever the trajectory is close to the GFP ($g,\lambda \ll
1$), not only in the GR regime. We may use them to derive a relation
between the coordinates $\bigl( g_{T}, \lambda_{T} \bigr)$ of the
trajectory's turning point $T$ at which it switches from decreasing to
increasing values of $\lambda$. (See Fig.\ \ref{fig4}.) Setting
$\partial_{t} \lambda =0$ in \eqref{25a} we obtain
\begin{align}
\lambda_{T} = \bigl( \varphi_{2} / 2 \pi \bigr) \, g_{T}.
\label{26-0}
\end{align}
The constant $\varphi_{2}$ is cutoff scheme--, i.\,e.\ 
$R^{(0)}$--dependent, but it is of order unity for any cutoff.
Therefore
\begin{align}
\lambda_{T} / g_{T} = \mathcal{O} (1).
\label{26-1}
\end{align}
Eqs.\ \eqref{26-0}, \eqref{26-1} are valid provided $g_{T},\lambda_{T}
\ll 1$. Later on we shall see that this is actually the case in
Nature.

After the trajectory has passed the turning point, $g$ keeps
decreasing and $\lambda$ increases. As a result, the second term on
the RHS of \eqref{25a}, $\varphi_{2} \, g / \pi$, gradually becomes
negligible compared to the first one, $- 2 \, \lambda$, so that the
flow equation becomes the canonical one. This marks the beginning of
the GR regime at $P_{1}$.

It is easy to solve the coupled differential equations \eqref{25}
exactly. They allow for two free constants of integration which we fix
by requiring $g \bigl( k_{T} \bigr) = g_{T}$ and $\lambda \bigl( k_{T}
\bigr) = \lambda_{T}$. By definition, $k_{T}$ is the scale at which the
trajectory passes through the turning point. The solution reads
\begin{subequations} \label{100}
\begin{align}
g (k) & = g_{T} \, \left( \frac{k}{k_{T}} \right)^{2}
\label{100a}
\\
\lambda (k) & = \frac{1}{2} \, \lambda_{T} \, \left( \frac{k_{T}}{k} \right)^{2}
\, \left[ 1 + \left( \frac{k}{k_{T}} \right)^{4} \right].
\label{100b}
\end{align}
\end{subequations}
The corresponding running of the dimensionful parameters is given by
\begin{subequations} \label{101}
\begin{align}
G (k) & = \frac{g_{T}}{k_{T}^{2}} = const
\label{101a}
\\
\Lambda (k) & = \frac{1}{2} \, \lambda_{T} \, k_{T}^{2}
\, \left[ 1 + \left( \frac{k}{k_{T}} \right)^{4} \right].
\label{101b}
\end{align}
\end{subequations}
Since the linear regime contains the GR regime, we may identify the
constant in \eqref{101a} with $G_{\text{lab}} \equiv
m_{\text{Pl}}^{-2}$. This entails the important relation $k_{T}^{2} =
g_{T} \, m_{\text{Pl}}^{2}$, or
\begin{align}
k_{T} = \sqrt{g_{T}\,} \, m_{\text{Pl}}.
\label{102}
\end{align}
We observe that a small $g_{T} \ll 1$ will lead to a large hierarchy
$m_{\text{Pl}} \gg k_{T}$.

We can use \eqref{102} in order to eliminate $k_{T}$ from \eqref{100}:
\begin{subequations} \label{103}
\begin{align}
g (k) & = \left( \frac{k}{m_{\text{Pl}}} \right)^{2}
\label{103a}
\\
\lambda (k) & = \frac{1}{2} \, g_{T} \lambda_{T} \,
\left( \frac{m_{\text{Pl}}}{k} \right)^{2} \,
\left[ 1 + \frac{k^{4}}{g_{T}^{2} \, m_{\text{Pl}}^{4}} \right].
\label{103b}
\end{align}
\end{subequations}
For later use we note that for any $k$ in the linear regime
\begin{align}
G (k) \, \Lambda (k) = g (k) \, \lambda (k) =
\frac{1}{2} \, g_{T} \lambda_{T} \,
\left[ 1 + \left( \frac{k}{k_{T}} \right)^{4} \right].
\label{104}
\end{align}

Looking at eqs.\ (\ref{101}a,b) we see that, while $G$ does not run at
all in the linear regime, the scale dependence of $\Lambda$ is
entirely due to the factor $\left[ 1 + \left( k / k_{T} \right)^{4}
\right]$. Once $k$ has become much smaller than $k_{T}$, after the
trajectory has ``turned left'', this factor approaches unity, and
$\Lambda$ effectively stops to run. By definition, this happens at
$P_{1}$, the starting point of the GR regime.

Let us make this statement more precise. Denoting by $k_{1}$ the scale
at which the trajectory passes through $P_{1}$, the requirement is
that $\left( k_{1} / k_{T} \right)^{4} \ll 1$. We quantify the
precision with which the $k^{4}$--term is negligible in the GR regime
by means of an exponent $\nu$. In terms of $\nu$, we define $k_{1}$
by
\begin{align}
k_{1} / k_{T} = 10^{-\nu}.
\label{105}
\end{align}
As a result, $\left( k / k_{T} \right)^{4}$ is smaller than
$10^{-4\nu}$ for all scales in the GR regime ($k < k_{1}$). A value
such as $\nu=1$ should be sufficient in practice. It makes sure that
in the GR regime $\Lambda$ is constant with a precision better than
$\unit[0.01]{\%}$.

For $k$ any scale in the GR regime we obtain from \eqref{103}, in
very good approximation,
\begin{subequations}
\begin{align}
g (k) & = \bigl( k / m_{\text{Pl}} \bigr)^{2}
\label{106a}
\\
\lambda (k) & = \tfrac{1}{2} \, g_{T} \lambda_{T} \,
\bigl( m_{\text{Pl}} / k \bigr)^{2}.
\label{106b}
\end{align} \label{106}
\end{subequations}
Similarly \eqref{101} yields the following constant values of the
dimensionful quantities:
\begin{subequations} \label{107}
\begin{align}
G (k) & = G_{\text{lab}}
\label{107a}
\\
\Lambda (k) & = \tfrac{1}{2} \, \lambda_{T} \, k_{T}^{2} = \tfrac{1}{2} \,
\Lambda \bigl( k_{T} \bigr).
\label{107b}
\end{align}
\end{subequations}
Remarkably, in the GR regime, $\Lambda$ differs from its value at the
turning point precisely by a factor of $1/2$. As a trivial
consequence,
\begin{subequations} \label{108}
\begin{align}
\left( G \Lambda \right) \Big \rvert_{\text{GR regime}} =
\tfrac{1}{2} \, \left( G \Lambda \right) \Big \rvert_{\text{turning point}}
\label{108a}
\end{align}
and \eqref{104} reduces to
\begin{align}
  G (k) \, \Lambda (k) = g (k) \, \lambda (k) = \tfrac{1}{2} \, g_{T}
  \lambda_{T}
\label{108b}
\end{align}
\end{subequations}
We shall come back to this relationship later on.

Next we derive various estimates for $g$ and $\lambda$ at $k_{T}$ and
$k_{1}$. If we evaluate \eqref{100} at $k = k_{1}$, neglecting the
$k^{4}$--term in \eqref{100b}, and use \eqref{105} we find\footnote{In
  order--of--magnitude equations such as \eqref{109} we suppress
  inessential factors of order unity.}
\begin{subequations} \label{109}
\begin{align}
g \bigl( k_{1} \bigr) & = g_{T} \, 10^{-2 \nu}
\label{109a}
\\
\lambda \bigl( k_{1} \bigr) & = \lambda_{T} \, 10^{+2 \nu}.
\label{109b}
\end{align}
\end{subequations}
Obviously $P_{1}$ has a $g$-- ($\lambda$--) coordinate which is
smaller (larger) than the corresponding coordinate of $T$ by a factor
$10^{-2\nu}$ ($10^{+2\nu}$), as it should be according to Fig.\ 
\ref{fig4}. Now we exploit eq.\ \eqref{26-1}. Neglecting factors of
order $1$ we have $g_{T} = \lambda_{T}$ which yields when combined
with \eqref{109}
\begin{align}
g \bigl( k_{1} \bigr) = 
\lambda \bigl( k_{1} \bigr) \, 10^{-4 \nu}.
\label{110}
\end{align}

Since $\lambda$ increases along the trajectory we know that
$\lambda_{T} < \lambda (k_{1}) < \lambda (k_{\text{lab}})$ where
$k_{\text{lab}}$ is any ``laboratory'' scale in the GR regime.
Therefore, as a consequence of the experimental result \eqref{24},
\begin{align}
  \lambda_{T} < \lambda \bigl( k_{1} \bigr) < \lambda \bigl(
  k_{\text{lab}} \bigr) \ll 1.
\label{111}
\end{align}
Since $\lambda_{T}$ and $g_{T}$ are almost equal, $g_{T}$ is small, too:
\begin{align}
g_{T} \approx \lambda_{T} \ll 1.
\label{112}
\end{align}
According to \eqref{110}, $g (k_{1})$ is even smaller than $\lambda
(k_{1})$. Hence, with $\lambda (k_{1}) \ll 1$ from \eqref{111}, it
follows that
\begin{align}
g \bigl( k_{1} \bigr) \ll 1, \quad\lambda \bigl( k_{1} \bigr) \ll 1.
\label{113} 
\end{align}

Eqs.\ \eqref{112} and \eqref{113} show that for the RG trajectory
realized in Nature the points $T$ and $P_{1}$ are located at an
extremely short distance to the GFP. The trajectory starts at the NGFP
with coordinates $g_{*}, \lambda_{*} = \mathcal{O} (0.1)$
\cite{oliver1,oliver2}. Then it follows the separatrix until, at very
tiny values of $g$ and $\lambda$, it gets ultimately driven away from
the GFP along its unstable $\lambda$--direction. In pictorial terms we
can say that the trajectory is squeezed deeply into the wedge formed
by the separatrix and the $g=0$--axis. As a consequence, it spends a
very long RG time near the GFP because the $\boldsymbol{\beta}$--functions are
small there. In this sense the RG trajectory which Nature has
selected is highly non--generic or ``unnatural''. It requires a precise
fine--tuning of the initial conditions, to be posed infinitesimally
close to the NGFP.
\subsection{Existence of a GR regime and the\\
cosmological constant problem}
\label{s3.3}
Why did Nature pick a trajectory which gets so ``unnaturally'' close
to the GFP? Why not, for example, one of those plotted in Fig.\ 
\ref{fig1} which always keep a distance of order unity to the GFP
and require no special fine--tuning? Of course questions of this kind
cannot be answered within QEG, for the same reason one
cannot compute the electron's charge or mass in QED.

However, it is fairly easy to show that a Universe based upon one of
the generic trajectories would look very different from the one we
know. The main difference is that, along a generic trajectory, no
sufficiently long GR regime would exist where classical gravity makes
sense at all. According to our previous discussion, the GR regime is
located in between a regime with strong UV renormalization effects
(spiraling around the NGFP related to asymptotic safety) and, most
probably, a second regime with a significant running of the parameters
in the IR. For classical GR to be applicable the UV and IR regimes
must be well separated. For a generic trajectory this is not the case,
however. The generic Einstein--Hilbert trajectories computed in
\cite{frank2} leave the UV regime at $k \approx m_{\text{Pl}}$ and
soon after they terminate at a $k_{\text{term}}$ not much smaller than
$m_{\text{Pl}}$; there is no GR regime which would last for a few
orders of magnitude at least.

The basic mechanism which allows for the emergence of a GR regime is
to fine--tune the trajectory in such a way that it spends a lot of RG
time near the GFP. In this manner the onset of the IR regime, within
the Einstein--Hilbert truncation characterized by the value of
$k_{\text{term}}$, gets enormously delayed, $k_{\text{term}}$ being
much smaller than $m_{\text{Pl}}$. If classical GR is correct up to a
length scale $L$, the trajectory must be such that its
$k_{\text{term}}^{-1}$ is larger than $L$ since at
$k_{\text{term}}^{-1}$ the IR effects are likely to become visible.

A quantitative estimate of $k_{\text{term}}$ can be obtained as
follows. In the linear regime the RG flow is explicitly given by eqs.\ 
\eqref{100}. Once $k$ is sufficiently low we leave this regime and the
full nonlinear Einstein--Hilbert flow equations have to be used. At
even smaller scales, the truncation breaks down and we should switch
to a more general one. Outside the linear regime, the
$\boldsymbol{\beta}$--functions are no longer small, hence the trajectory has a
comparatively high ``speed'' there. As a result, it takes the
trajectory much longer to go from $T$ to the boundary of the linear,
i.\,e.\ GR regime than from there to a point close to $\lambda = 1/2$.
Therefore we may use eq.\ \eqref{106b} for $\lambda (k)$ in the GR
regime in order to derive an estimate for $k_{\text{term}}$. We
identify $k_{\text{term}}$ with the scale where, according to
\eqref{106b}, the value $\lambda = 1/2$ is reached: $g_{T} \lambda_{T}
\, \left( m_{\text{Pl}} / k_{\text{term}} \right)^{2} =1$. In this
rough approximation,
\begin{subequations} \label{114}
\begin{align}
k_{\text{term}} 
& = \sqrt{g_{T} \lambda_{T} \,} \, m_{\text{Pl}}
\label{114a}
\\
& = \bigl( \varphi_{2} / 2 \pi \bigr)^{1/2} \, g_{T} \,  m_{\text{Pl}}
\label{114b}
\end{align}
\end{subequations}
or, in terms of length scales,
\begin{align}
k_{\text{term}}^{-1} = \frac{\ell_{\text{Pl}}}{\sqrt{g_{T} \lambda_{T}\,}\,}.
\label{115}
\end{align}
This equation shows explicitly that by making $g_{T} \approx
\lambda_{T}$ small, $k_{\text{term}}^{-1}$ can be made as large as we
like.

Thus we have demonstrated that \textit{only a ``non--generic''
  trajectory with an ``unnaturally'' small $g_{T} \approx \lambda_{T}
  \ll 1$ does give rise to a long GR regime comprising many orders of
  magnitude}.

In \eqref{102} we saw that the turning point is passed at $k_{T} =
\sqrt{g_{T}\,} \, m_{\text{Pl}}$. Ignoring the $\mathcal{O}
(1)$--factor $\left( \varphi_{2} / 2 \pi \right)$ in \eqref{114b} this
entails $k_{\text{term}} = \sqrt{g_{T}\,} \, k_{T}$. As a result,
there exists an exactly symmetric ``double hierarchy'' among the three
mass scales $k_{\text{term}}$, $k_{T}$, and $m_{\text{Pl}}$:
\begin{align}
\frac{k_{\text{term}}}{k_{T}} = \sqrt{g_{T}\,} \ll 1, \qquad
\frac{k_{T}}{m_{\text{Pl}}} = \sqrt{g_{T}\,} \ll 1.
\label{116}
\end{align}
Therefore, on a logarithmic scale, $k_{T}$ is precisely in the middle
between $m_{\text{Pl}}$ and $k_{\text{term}}$. Thanks to the smallness
of $g_{T}$ it is many orders of magnitude away from either end.

The emergence of a long regime where gravity is essentially classical
is one of the benefits we get from the unnaturalness of the trajectory
chosen by Nature. Another one is that, in this classical regime, the
cosmological constant is automatically small. Inserting \eqref{102}
into \eqref{107b} we obtain the following $\Lambda$ in the GR regime:
\begin{align}
\Lambda (k) = \tfrac{1}{2} \, g_{T} \lambda_{T} \, m_{\text{Pl}}^{2} = const.
\label{117}
\end{align}
Again thanks to the smallness of $g_{T}$ and $\lambda_{T}$, the
cosmological constant is much smaller than $m_{\text{Pl}}^{2}$. Up to
a factor of order unity,
\begin{align}
\frac{\Lambda}{m_{\text{Pl}}^{2}} \Bigg \rvert_{\text{GR regime}} =
g_{T}^{2} \ll 1.
\label{118}
\end{align}
Thus we may conclude that the very fine--tuning which gives rise to a long GR
regime at the same time implies a large hierarchy between the $\Lambda$ in
this GR regime and $m_{\text{Pl}}^{2}$, which often is considered its
``natural'' value.

This observation provides a solution to the ``cosmological constant
problem'' by realizing that it is actually part of a much more general
naturalness problem. Rather than ``Why is $\Lambda$ so small?'' the
new question is ``Why does gravity behave classically over such a long
interval of scales?''.

The basic reason for this connection is very simple. Denoting the
cosmological constant in the GR regime by $\Lambda (k_{\text{lab}})
\equiv \Lambda_{\text{lab}}$ we have $\lambda (k) =
\Lambda_{\text{lab}} / k^{2}$ there. This $\lambda (k)$ approaches
unity so that the IR renormalization effects become strong once $k$ is
of the order $\sqrt{\Lambda_{\text{lab}}\,}$. Hence, roughly,
\begin{align}
k_{\text{term}} \approx \sqrt{\Lambda_{\text{lab}}\,}
\label{119}
\end{align}
which shows that $k_{\text{term}}$ is small if, and only if,
$\Lambda_{\text{lab}}$ is small.

This is precisely what one finds by numerically solving the flow
equations: There do not exist any Type IIIa trajectories which, on the
one hand, admit a long classical regime, and on the other hand, have a
large cosmological constant. As a consequence, \textit{the smallness
  of the cosmological constant poses no naturalness problem beyond the
  one related to the very existence of a classical regime in the
  Universe}.

Strictly speaking this resolution of the cosmological constant problem
is only a partial one for the following reasons. (1) We analyzed the
flow equations of pure gravity only, but we believe that the inclusion
of matter fields (forming symmetry breaking condensates, etc.) will
not change the general picture. Since anyhow only massless particles
contribute to the $\boldsymbol{\beta}$--functions for $k \to 0$, it is hardly
possible that the matter fields destroy the IR renormalizations near
$\lambda = 1/2$. But if they survive the estimate \eqref{119} remains
intact, again implying that a long GR regime requires a small
$\Lambda_{\text{lab}}$. (2) Our argument refers to
$\Lambda_{\text{lab}}$ rather than the cosmologically relevant
$\Lambda (H_{0})$. Because of the IR effects, $\Lambda_{\text{lab}}$
and $\Lambda (H_{0})$ differ in principle, but we shall see later on
that in Nature this difference is small compared to the notorious 120
orders of magnitude one has to cope with.

In fact, the upper boundary of the linear regime is roughly the Planck
scale since, according to \eqref{103}, $g (k)$ and $\lambda (k)$ are
of the order of $g_{*}$ and $\lambda_{*}$ slightly below $k =
m_{\text{Pl}}$. Between this scale and the turning point the change of
$\Lambda$ is enormous. From \eqref{101b} with \eqref{102} and
\eqref{112} we obtain $\Lambda (m_{\text{Pl}}) = \tfrac{1}{2} \,
\Lambda (k_{T}) / g_{T}^{2} \gg \Lambda (k_{T})$. In the next section
we shall see that $g_{T} \approx 10^{-60}$, whence $\Lambda (k_{T})
\approx 10^{-120} \, \Lambda (m_{\text{Pl}})$.
\subsection{Is the Hubble scale within the classical regime?}
\label{s3.4}
The observational (CMBR, supernova, etc.) data, when interpreted
within standard cosmology, show that the present vacuum energy density
of the Universe is very close to the critical one, implying that
$\Lambda$ is of the order of $H_{0}^{2}$. (The general relationship is
$\Lambda = 3 \, \Omega_{\Lambda} \, H_{0}^{2}$, and the data favor
$\Omega_{\Lambda} \approx 0.7$).

We can now ask whether, if $\Lambda$ is as large as $H_{0}^{2}$, the
Hubble scale is still within the GR regime. If we interpret
$H_{0}^{2}$ as a ``laboratory'' value, $\Lambda_{\text{lab}} \approx
H_{0}^{2}$, the estimate \eqref{119} yields
\begin{align}
k_{\text{term}} \approx H_{0}.
\label{120}
\end{align}
This is a very remarkable and intriguing result: \textit{On the RG
  trajectory Nature has picked, the Hubble scale $k = H_{0}$ is
  precisely at the boundary of the GR regime.} At distances small
compared to the Hubble length $\ell_{H} \equiv H_{0}^{-1}$ classical
GR is a good approximation, but on length scales $\ell \gtrsim
\ell_{H}$ the IR renormalization effects become important so that its
use is questionable there.

Since $k = H_{0}$ is just at the boundary of the GR regime it is
plausible to assume that the cosmological values $G (H_{0})$ and
$\Lambda (H_{0})$ do not differ too much from $G_{\text{lab}}$ and
$\Lambda_{\text{lab}}$, respectively. (Later on we shall see that the
difference should not be more than one or two orders of magnitude.) If
so, we are in a position to completely fix the parameters of the RG
trajectory because we can fit the measured $\Lambda$--value to the
trajectories of the linearized flow. Inserting $k = H_{0}$, $\Lambda
(k) = H_{0}^{2}$, and $G (k) = m_{\text{Pl}}^{-2}$ into \eqref{108b}
we find
\begin{align}
\tfrac{1}{2} \, g_{T} \lambda_{T} = g (k) \, \lambda (k) =
\bigl( H_{0} / m_{\text{Pl}} \bigr)^{2}.
\label{121}
\end{align}
Since $g_{T}$ and $\lambda_{T}$ are approximately equal, and since the
``experimental'' value for $H_{0} / m_{\text{Pl}}$ is known to be
about $10^{-60}$, eq.\ \eqref{119} allows for an explicit
determination of $g_{T}$ and $\lambda_{T}$:
\begin{align}
g_{T} \approx \lambda_{T} \approx H_{0} / m_{\text{Pl}}
\approx 10^{-60}.
\label{26-4}
\end{align}
This number is indeed extremely small but, consistently with
the picture drawn so far, still larger than $g (k_{\text{lab}})
\lessapprox 10^{-70}$.

With \eqref{26-4} the ``double hierarchy'' \eqref{116} comprises 30
orders of magnitude between $k_{\text{term}}$ and $k_{T}$, and between
$k_{T}$ and $m_{\text{Pl}}$, respectively. Combining eq.\ \eqref{114a}
for $k_{\text{term}}$ with \eqref{121} we rediscover that
$k_{\text{term}} \approx H_{0}$, and using eq.\ \eqref{102} we obtain
the scale at the turning point:
\begin{align}
  k_{T} \approx \sqrt{H_{0} \, m_{\text{Pl}} \,} \approx 10^{-30} \,
  m_{\text{Pl}}.
\label{122} 
\end{align}
The associated length scale is
\begin{align}
k_{T}^{-1} \approx 10^{30} \, \ell_{\text{Pl}} \approx \unit[10^{-3}]{cm}.
\label{123}
\end{align}
It is very exciting that this is a truly macroscopic length scale,
very far away from the Planck regime which is usually thought to be
the realm of the quantum gravitational UV renormalization effects.
Note that the ``turning left'' of the trajectory, despite the large
value of \eqref{123}, is still an effect of the ``UV type'' in the
sense that it has nothing to do with the IR phenomena occurring when
$\lambda \approx 1/2$. It happens so extremely late because of
the enormous fine--tuning of the trajectory.

Allowing a margin of one order of magnitude ($\nu=1$), say,
\eqref{123} implies a beginning of the GR regime at
\begin{align}
k_{1}^{-1} \approx \unit[10^{-2}]{cm}.
\label{124}
\end{align}
At length scales $\ell \equiv k^{-1}$ larger than this value $\Lambda$
is constant with high precision. But for $\ell \ll k_{1}^{-1}$ it has
a strong scale dependence $\propto k^{4}$ given by eq.\ \eqref{101b}.
Since $\lambda_{T} \, k_{T}^{2} / 2 = \Lambda_{\text{lab}} \approx
H_{0}^{2}$ we may rewrite this equation in the form
\begin{align}
\Lambda (k) \approx H_{0}^{2} \, \left[ 1 + \bigl( k / k_{T} \bigr)^{4} 
\right].
\label{125}
\end{align}

Should we then expect to see strong violations of GR at the millimeter
or micrometer scale? The answer is probably no. While the variation of
$\Lambda (k)$ is indeed very strong, between $k^{-1} = \unit[1]{\mu
  m}$ and $k^{-1} = \unit[1]{nm}$, say, it changes by 12 orders of
magnitude, the absolute value of $\Lambda$ is proportional to
$H_{0}^{2}$ which is extremely tiny according to all standards of
terrestrial experiments, of course. Because of the practical problems
with measuring $\Lambda$ at non--cosmological distances the running of
$\Lambda$ could possibly remain undetected over many orders of magnitude.

By combining \eqref{106} with \eqref{121} we obtain $g$ and $\lambda$
at the Hubble scale:
\begin{align}
g \bigl( H_{0} \bigr) \approx 10^{-120}, \quad
\lambda \bigl( H_{0} \bigr) = \mathcal{O} (1).
\label{126}
\end{align}
This approximation neglects the IR effects close to $H_{0}$, but
should be correct as far as the orders of magnitude are concerned.
%
%
%
%
%
%
%
%
\section{The RG trajectory realized in Nature: the deep IR}
\label{s4}
\subsection{Leaving the GR regime}
\label{s4.1}
Denoting the scales corresponding to the boundary points of the GR
regime, $P_{1}$ and $P_{2}$, by $k_{1}$ and $k_{2}$, respectively, the
general picture of the trajectory which Nature has chosen can be
summarized as follows. For $k = \infty$ it starts infinitesimally
close to the NGFP, runs very close to the separatrix until, at
$k_{T}$, it turns left in the very last moment before hitting the GFP,
at $g_{T}, \lambda_{T} \ll 1$, then enters the GR regime at the scale
$k_{1}$ and leaves it at $k_{2}$. In the entire GR regime $G$ and
$\Lambda$ are constant, and $\lambda (k) \ll 1$.

At scales $k < k_{2}$ immediately below the GR regime there exists a
regime where the renormalization effects become appreciable, but are
still weak enough for the Einstein--Hilbert truncation to be reliable.
The mechanism discussed in Section \ref{s2} causes $G (k)$ to increase
with decreasing $k$. The dimensionless cosmological constant $\lambda$
is still much smaller than $1/2$ there.

At even smaller scales, $\lambda$ approaches $1/2$, and the
Einstein--Hilbert truncation becomes unreliable as it would yield an
unbounded growth of $G (k)$ and $\eta_{\text{N}} \to - \infty$,
$\text{d} G / \text{d} k \to - \infty$ at a nonzero $k_{\text{term}}$.

It is our main hypothesis that exact QEG or a sufficiently general
truncation ``tames'' this divergent behavior and leads to a bounded
growth of $G (k)$, i.\,e.\ a finite $\eta_{\text{N}}$, along a RG
trajectory which passes smoothly through $k_{\text{term}}$ and can be
continued to $k=0$. We believe that those IR renormalization effects
lead to observable physical effects, and that the scale at which they
occur is approximately given by the value of $k_{\text{term}}$
obtained from the Einstein--Hilbert truncation.

The truncations needed for a reliable ab initio computation of the IR
effects within QEG are much more complicated than those used so far.
They contain many more couplings beyond $g$ and $\lambda$. The
resulting RG trajectory can be visualized as a curve in a high
dimensional ``theory space'' containing the $g$-$\lambda$--plane as a
subspace. In general one has to distinguish the projection of the
exact trajectory onto the $g$-$\lambda$--plane and the trajectory
implied by the Einstein--Hilbert truncation whose theory space consists
of this plane only. As long as this truncation is reliable the two
curves in the $g$-$\lambda$--plane do not differ much, but the
deviations become large when the trajectory runs out of the domain of
validity of the truncation. For the RG trajectory realized in Nature
this is the case near the question mark in Fig.\ \ref{fig4}, close to
$\lambda = 1/2$.

Clearly it would be desirable to employ such improved truncations
which allow for a continuation of the Type IIIa trajectories to $k=0$,
but because of the enormous mathematical complexity of the resulting
flow equations this seems to be out of reach using the presently
available computational techniques. The situation is similar to QCD
where the analogous nonperturbative IR phenomena such as color
confinement or chiral symmetry breaking are notoriously difficult to
deal with analytically.

It is of crucial importance to find out at which scale the IR effects
become strong. In principle $k_{\text{term}}$ can be determined by
measuring $G (k_{\text{lab}})$ and $\Lambda (k_{\text{lab}})$ in the
``laboratory'' and integrating the Einstein--Hilbert flow equations
towards smaller $k$ until one reaches $\lambda = 1/2$ at
$k_{\text{term}}$. However, in practice this does not work because it
is very hard to measure $\Lambda$ in a non--cosmological
``laboratory''.

A first hint about the value of $k_{\text{term}}$ comes from the
argument in Subsection \ref{s3.4}. Given the measured value of
$\Lambda$, $k_{\text{term}}$ is of the order of the Hubble scale. We
take this as a strong indication suggesting that at least in cosmology
the IR effects are ``at work'' already. However, in order to assess
their relevance at galactic scales, a more precise estimate is needed.

Since we have no analytical tools yet to investigate those IR effects
we can resort to the following phenomenological strategy for obtaining
information about the RG trajectory: We make an ansatz for the
trajectory, work out its consequences, and compare them to the
observational data.\footnote{In this context it is important to
  understand that the RG trajectory is a universal object which is not
  specific to any particular system: $\Gamma_{k} [g_{\mu \nu}]$
  defines an effective field theory for \textit{all} systems whose
  typical scale is of the order of $k$. The specific features of the
  system under consideration enter only when it comes to identifying
  $k$ in terms of dynamical or geometrical data, see \cite{cosmo1,h1,h2}.}
The so--called IR fixed point model developed in \cite{cosmo2,elo} is
an attempt in precisely this direction. In the next subsection we
describe its status in the context of the present discussion.
\subsection{The IR fixed point model}
\label{4.2}
The IR fixed point model \cite{cosmo2,elo} is a cosmology based upon
the hypothesis that, for $k \to 0$, the projected trajectory $\bigl( g
(k), \lambda (k) \bigr)$ runs into a non--Gaussian IR fixed point
$\bigl( g_{*}^{\text{IR}}, \lambda_{*}^{\text{IR}} \bigr)$, different
from the UV fixed point it emanates from. This assumption implies that
in the deep IR the dimensionful parameters run as
\begin{align}
G (k) = g_{*}^{\text{IR}} / k^{2}, \quad
\Lambda (k) = \lambda_{*}^{\text{IR}} \, k^{2}.
\label{27}
\end{align}
The behavior of $\Lambda (k)$ is the same as in \eqref{19-2} which was
motivated by the analogy with the scalar theory, and $G (k) \propto 1
/ k^{2}$ is a plausible ansatz for the growth of Newton's constant.
Since $\boldsymbol{\beta}_{g} = \left( 2 + \eta_{\text{N}} \right) \,
g$ holds true in general, but not necessarily with the
$\eta_{\text{N}}$ of \eqref{11}, the anomalous dimension at the fixed
point is $\eta_{\text{N}} = -2$.

Note that while the IR renormalization according to \eqref{27} has a
strong effect on $G (k)$ and $\Lambda (k)$ separately, it leaves the
product $ G (k) \, \Lambda (k)$ invariant: $\Lambda (k)$ decreases at
the same rate $G (k)$ increases towards the IR.

In a homogeneous and isotropic Universe the relevant cutoff scale is
$k = \widehat \xi / t$ to leading order\footnote{We consider only
  power law solutions for which $\widehat \xi / t$ is proportional to
  the Hubble parameter $H (t)$.} \cite{cosmo1,cosmo2}. Here $t$ is the
cosmological time and $\widehat \xi$ is a constant of order unity.
This ``cutoff identification'' turns $G (k)$ and $\Lambda (k)$ into
functions of time: $G (t) \equiv G (k = \widehat \xi / t)$, and
similarly for $\Lambda$. Replacing the constants $G$ and $\Lambda$ in
the classical Einstein equation with $G (t)$ and $\Lambda (t)$ one
obtains a ``RG improved'' field equation whose Robertson--Walker--type
solutions can be studied \cite{cosmo1,cosmo2}. For the $k$--dependence
\eqref{27} one finds an accelerating Universe with a scale factor $a
(t) \propto t^{4/3}$. The fixed point not only forces the Universe to
enter an epoch of accelerated expansion for $t \to \infty$, it also
explains without any fine--tuning of parameters why in the late
Universe the matter energy density equals approximately the vacuum
energy density, thus providing a natural solution to the ``cosmic
coincidence puzzle''. Confronting the infrared fixed point model with
the observational data (supernovae, compact radio sources, CMBR, etc.)
it performs as well as the best--fit Friedmann model. (See ref.\ 
\cite{elo} for further details.)

The fixed point solution to the improved Einstein equation exists only
if $\lambda_{*}^{\text{IR}}$ and $\widehat \xi$ are such that
$\lambda_{*}^{\text{IR}} \, {\widehat \xi}^{~2} = 8/3$, implying that
$\lambda_{*}^{\text{IR}}$ has to be of order unity since $\widehat \xi
= \mathcal{O} (1)$. Furthermore, $g_{*}^{\text{IR}}$ is found to be
of the order $\left[ G (t_{0}) / G_{\text{lab}} \right] \, \left(
  H_{0} / m_{\text{Pl}} \right)^{2}$. Here $G (t_{0})$ denotes the
present Newton constant at cosmological scales. According to the
analysis of ref.\ \cite{elo}, the observational data imply that the
ratio $G (t_{0}) / G_{\text{lab}}$ can comprise at most one or two
orders of magnitude (in any case a number by far smaller than 120, say).
Therefore, using $H_{0} / m_{\text{Pl}} \approx 10^{-60}$, we find
that the model is consistent only provided
\begin{align}
g_{*}^{\text{IR}} \approx 10^{-120}, \quad
\lambda_{*}^{\text{IR}} \approx \mathcal{O} (1).
\label{28}
\end{align}

This model is based upon the assumption that the IR effects lead to
the formation of a fixed point into which the trajectory is attracted
for $k \to 0$. The solubility of the improved Einstein
equation and the observational data then require that the projection
of the fixed point onto the $g$-$\lambda$--plane is located at the
coordinates \eqref{28}. The comparison with \eqref{126} shows that the
point $\bigl( g_{*}^{\text{IR}}, \lambda_{*}^{\text{IR}} \bigr)$ is
precisely in the region where the Einstein--Hilbert approximation
breaks down and something new must happen. (Near the question mark in
Fig.\ \ref{fig4}.)

This consistency is an important success of the fixed point model. It
describes the simplest possible behavior of the (projected) RG
trajectory in the deep IR: the strong quantum corrections near
$\lambda = 1/2$ simply bring the running of $g (k)$ and $\lambda (k)$
to a complete standstill, at $ g_{*}^{\text{IR}}$ and
$\lambda_{*}^{\text{IR}}$, respectively.

We can think of the cosmological time evolution as an evolution of the
Universe along the RG trajectory with $k$ corresponding to $1/t$ or $H
(t)$, which is essentially the same here. As long as the Universe
stays in the GR regime the classical Friedmann equations are a valid
description, but as soon as $\lambda$ gets close to $1/2$ the fixed
point cosmology takes over, the Universe starts accelerating, and the
vacuum and matter energy densities become approximately equal
($\Omega_{\Lambda} \approx \Omega_{\text{M}}$). These predictions are
in remarkable agreement with what we learned about the Universe
from the astrophysical data which became available during the past few
years. (See refs.\ \cite{cosmo2,elo} for further details.)

Even if the finer details of the IR fixed point model are perhaps not
completely correct quantitatively it demonstrates that an increase
(decrease) of $G$ ($\Lambda$) near the Hubble scale would explain many
of the observed features of the late Universe in an extremely natural
way. In a sense, the renormalization effects would mimic the presence
of ``quintessence'' which was invented in order to explain these
features.
\subsection{Where does the IR running set in?}
\label{s4.3}
Motivated by the estimate $k_{\text{term}} \approx H_{0}$ from
Subsection \ref{s3.4} and the phenomenological success of the IR fixed
point model we believe that at the present Hubble scale the IR effects
have an observable and qualitatively important impact on gravitational
physics already. Now the crucial question is at which scale
$k_{1} > H_{0}$ precisely the GR regime ends and the parameters start
running.

If we knew the exact RG trajectory we could set up the corresponding
RG improved Friedmann equations, interpret the observational data
measured in the late Universe within this framework, and deduce
$k_{1}$ from the redshift of the epoch in which deviations from
classical cosmology become visible. In ref.\ \cite{elo} this analysis
was performed within the fixed point model but the statistical quality
of the presently available data (on high redshift type Ia
supernovae) is still too poor to allow for a precise
determination of $k_{1}$. It became fairly clear, however, that
between $k=k_{1}$ and $k=H_{0}$ Newton's constant cannot have changed
by more than 1 or 2 orders of magnitude.

Thus we are led to look for possible manifestations of the IR running
in gravitationally bound structures which are small on cosmological
scales. Clearly the natural place to search for such phenomena are
galactic systems. One of their most striking features, as to yet
completely unexplained, is the discrepancy between their directly
observable mass (due to ``luminous matter'') and the mass inferred from
the observed motions. This mass discrepancy occurs in all types of
galactic systems, from dwarf galaxies to superclusters. Its magnitude
ranges from a factor of a few at the kiloparsec scale to a factor of a
few hundred at the megaparsec scale \cite{padman}. The most popular
attempt at explaining this discrepancy is the dark matter hypothesis.

In the present paper we propose that the mass discrepancy is actually
not due to the presence of dark matter but rather to the IR
renormalization effects. Taking this hypothesis seriously, we shall
try to learn something about the RG trajectory by taking advantage of
what is known about galaxies. There is a large amount of
observational data \cite{combbook} showing that the orbits of ``test
particles'' moving in typical spiral galaxies cannot be explained
by Newtonian gravity if the gravitational attraction is due to the
luminous matter only. Usually the rotation curve $v (r)$, the velocity
on a circular orbit as a function of the distance to the center,
becomes approximately constant for large distances $r$, in
contradiction to Kepler's law.

In the following sections we shall demonstrate, by means of an appropriate
renormalization group improvement, that such violations of
Kepler's law are exactly what one would expect to occur if $G
(k)$ shows a certain scale dependence at galactic scales already.

A first argument demonstrating that this scenario is not completely
unreasonable is the following. A mass discrepancy of a factor of a few
hundred at the cluster scale indicates that, very roughly, $G$ at
those scales should not differ from $G_{\text{lab}}$ by more than a
similar factor of a few hundred. This is nicely consistent with what
we got for $G (t_{0}) / G_{\text{lab}}$ in the fixed point model
\cite{elo} at somewhat larger scales. Given the many orders of
magnitude we are dealing with here this consistency is certainly
nontrivial.

A comparatively small ratio $G_{\text{galaxy}} / G_{\text{lab}}$ of
$\mathcal{O} (10)$ or perhaps $\mathcal{O} (100)$, and a similar ratio
in cosmology, justifies our earlier use of the linearized flow
equation down to $k = H_{0}$ which neglects the running between the
end of the GR regime and the Hubble scale.

While a renormalization of $G$ and $\Lambda$ by a factor of 10 is
extremely little compared to the many orders of magnitude their values
changed in the UV, it leads to significant modifications of classical
Newtonian gravity ``in the sky'', at astrophysical scales.
\subsection{Running $\boldsymbol{G}$ at galactic scales}
\label{s4.4}
Henceforth we shall assume that the IR running starts somewhere
between solar system and galactic scales.\footnote{As for the QCD
  analogy, a feature of QEG not shared by QCD is the long intermediate
  classical regime, as a result of which deviations from the
  $1/r$--potential occur near $k_{\text{term}} \lll m_{\text{Pl}}$
  only, while in QCD this happens at the scale $\Lambda_{\text{QCD}}$
  which is analogous to $m_{\text{Pl}}$ in other respects.}
Concentrating on the running of Newton's constant\footnote{For small
  isolated systems such as an individual galaxy the cosmological
  constant plays no important role probably. (Cosmological effects
  will be neglected in our discussion of galaxies.)} and, for
technical simplicity, spherically symmetric model galaxies, the first
step towards their RG improved dynamics is as follows. Given the
function $G = G (k)$ pertaining to the RG trajectory realized in
Nature we convert its $k$--dependence to a distance dependence by
means of an appropriate cutoff identification \cite{bh}. As long as
curvature effects are small, the obvious choice for spherically
symmetric systems is $k = \xi / r$, with $\xi$ a constant of order
unity \cite{h2}. We define the position dependent Newton constant as
\begin{align}
G (r) \equiv G ( k = \xi / r).
\label{29}
\end{align}
By the very construction of the effective average action
\cite{avactrev}, $G (r)$ is the parameter which appears in the effective
field equations for quantities averaged over a volume of linear
dimension $\approx r$.

Using $G (r)$ we can try to RG improve the classical Newton potential
$-G_{\text{lab}} M / r$ by substituting $G_{\text{lab}} \to G (r)$. In
the case of the UV renormalization effects \cite{mr} it is known that
this procedure reproduces the results of the explicit perturbative
calculation \cite{donoghue,bohr}. However, as we shall discuss in the
next section the improved action approach predicts a
``nonperturbative'' large distance correction which is even more
important than the one taken into account by replacing $G_{\text{lab}}
\to G (r)$ in the potential.

The most important question is which $k$--dependence we should expect
at a typical galactic scale of $k^{-1} = \unit[1]{kpc} \cdots
\unit[100]{kpc}$, say. As it will turn out, for
the trajectory realized in Nature this scale is already outside the
domain of validity of the nonlinear Einstein--Hilbert flow equations.
Therefore, the best we can do is to make an ansatz for $G (k)$ which
reproduces the ``phenomenology'' of galaxies as well as possible, and
at the same time provides (part of) a natural interpolation between
the GR regime and the IR fixed point behavior.

Classical GR is characterized by a vanishing anomalous dimension,
$\eta_{\text{N}} =0$. Switching on the renormalization effects, the
theoretically simplest option is to assume a nonzero, but constant
$\eta_{\text{N}}$. Since $\partial_{t} G = \eta_{\text{N}} \, G$ this
entails a power law $G (k) \propto k^{\eta_{\text{N}}}$. As $G (k)$ is
an increasing function of $k$, $\eta_{\text{N}}$ is negative. We shall
set $\eta_{\text{N}} = -q$ therefore with a positive constant $q$.
Thus we tentatively assume that
\begin{align}
G (k) \propto 1 / k^{q}
\label{30}
\end{align}
for the range of $k$--values which are relevant to the structure of
galaxies. Note that the assumption of an approximately constant
$\eta_{\text{N}}$ is not too restrictive since this range comprises
only 2 or 3 orders of magnitude, from $\unit[1]{kpc}$ and $100$ or
$\unit[1000]{kpc}$, say. The power law \eqref{30} leads to the radial
dependence
\begin{align}
G (r) \propto r^{q}.
\label{31}
\end{align}
The consequences of this $r$--dependence we are going to explore in a
moment. It will lead to phenomenologically acceptable rotation curves
if $q \approx 10^{-6}$.

There exists an intriguing scenario which would give rise to an
approximately constant $\eta_{\text{N}}$ during some $k$--interval in
a natural way. Let us assume that, at galactic scales, the RG
trajectory approaches a further NGFP, different from the ones
discussed above, spends some time in its vicinity, and is finally
driven away from it.  This fixed point $\bigl( g_{*}, \lambda_{*},
\cdots \bigr)$ must have the $g$--coordinate $g_{*} =0$ because then
$\boldsymbol{\beta}_{g} \equiv \left( 2 + \eta_{\text{N}} \right) \,
g$ vanishes there without constraining the value of $\eta_{\text{N}}$.
(Every NGFP with $g_{*} \neq 0$ has $\eta_{*}=-2$.) Hence, if
$\eta_{*} \equiv \eta_{\text{N}} (g_{*} =0, \lambda_{*}, \cdots)
\equiv -q$ is nonzero, the smallness of the
$\boldsymbol{\beta}$--functions near the NGFP implies that
$\eta_{\text{N}} \bigl( g (k), \lambda (k), \cdots \bigr)$ is nonzero
and approximately $k$--independent there.

A fixed point of this kind could only arise within truncations which
are much more general than the ones used so far, as the result of a
strongly nonperturbative dynamics. (Of course, the perturbative
quantization of the Einstein--Hilbert action yields $\eta_{\text{N}}=0$
at $g=0$, to all orders.)

The projection of the new fixed point onto the $g$-$\lambda$--plane of
Fig.\ \ref{fig4} is located exactly on the horizontal ($g=0$) axis.
What makes this scenario particularly plausible is that we know
already that the RG trajectory realized in Nature does indeed get
extremely close to the $g=0$--axis, as it would be necessary to make
$\eta_{\text{N}}$ approximately $k$--independent. However, even if
this additional NGFP should not exist as a strict fixed point it is
still quite plausible that a behavior similar to \eqref{30} can
prevail for a short RG time and this is all we need.
%
%
%
%
%
%
%
%
\section{Galaxy rotation curves}
\label{s5}
In this section we apply the improved--action approach to the problem
of the flat galaxy rotation curves. For further details about the
general approach we refer to \cite{h1}, and to \cite{h2} for a
detailed analysis of spherically symmetric, static ``model galaxies''.

The idea is to start from the classical Einstein--Hilbert action
$S_{\text{EH}} = \int \!\! \text{d}^{4} x~\sqrt{-g\,} \,
\mathscr{L}_{\text{EH}}$ with the Lagrangian $\mathscr{L}_{\text{EH}}
= \left( R - 2 \, \Lambda \right) / \left( 16 \pi \, G \right)$ and to
promote $G$ and $\Lambda$ to scalar fields. This
leads to the modified Einstein--Hilbert (mEH) action
\begin{align}
S_{\text{mEH}} [g,G,\Lambda] =
\frac{1}{16 \pi} \, \int \!\! \text{d}^{4} x~
\sqrt{-g\,} \bigg\{
\frac{R}{G (x)} - 2 \, \frac{\Lambda (x)}{G (x)}
\bigg\}.
\label{40}
\end{align}
The resulting theory has certain features in common with Brans--Dicke
theory; the main difference is that $G (x)$ (and $\Lambda (x)$) is a
prescribed ``background field'' rather than a Klein--Gordon scalar as
usually. Upon adding a matter contribution the action \eqref{40}
implies the modified Einstein equation\footnote{In \cite{h1} and
  \cite{h2} a further contribution, $\theta_{\mu \nu}$, was added to
  the energy--momentum tensor in order to describe the 4--momentum of
  the field $G (x)$. Its form is not completely fixed by general
  principles. But as it does not affect the Newtonian limit \cite{h2}
  we set $\theta_{\mu \nu} \equiv 0$ here.}
\begin{align}
G_{\mu \nu} = - \Lambda (x) \, g_{\mu \nu}
+ 8 \pi \, G (x) \, \bigl( T_{\mu \nu} + \Delta T_{\mu \nu} \bigr).
\label{41}
\end{align}
Here $\Delta T_{\mu \nu}$ is an additional contribution to the
energy--momentum tensor due to the $x$--dependence of $G$:
\begin{align}
\Delta T_{\mu \nu} \equiv \frac{1}{8 \pi} \,
\bigl(
D_{\mu} D_{\nu} - g_{\mu \nu} \, D^{2}
\bigr) \, \frac{1}{G (x)}.
\label{42}
\end{align}
The field equation \eqref{41} is mathematically consistent provided
$\Lambda (x)$ and $G (x)$ satisfy a ``consistency condition'' which
insures that the RHS of \eqref{41} has a vanishing covariant
divergence.

In ref.\ \cite{h2} we analyzed the weak field, slow--motion
approximation of this theory for a time--independent Newton constant $G
= G (\mathbf{x})$ and $\Lambda \equiv 0$. In this (modified) Newtonian
limit the equation of motion for massive test particles has the usual
form, $\ddot {\mathbf{x}} (t) = - \nabla \phi$, but the potential $\phi$
obeys a modified Poisson equation,
\begin{subequations} \label{43}
\begin{align}
\nabla^{2} \phi = 4 \pi \, \overline{G} \, \rho_{\text{eff}}
\label{43a}
\end{align}
with the effective density\footnote{In this paper we neglect in
  $\rho_{\text{eff}}$ an inessential term $\mathcal{N} \, \rho$
  relative to $\rho$.}
\begin{align}
\rho_{\text{eff}} = \rho + \bigl( 8 \pi \, \overline{G} \, \bigr)^{-1} \,
\nabla^{2} \mathcal{N}.
\label{43b} 
\end{align}
\end{subequations}
In deriving \eqref{43} it was assumed that $T_{\mu \nu}$ describes
pressureless dust of density $\rho$ and that $G (\mathbf{x})$ does not
differ much from the constant $\overline{G}$. We use the
parameterization
\begin{align}
G (\mathbf{x}) = \overline{G} \, \bigl[ 1 + \mathcal{N} (\mathbf{x}) \bigr]
\label{44}
\end{align}
and assume that $\mathcal{N} (\mathbf{x}) \ll 1$. More precisely, the
assumptions leading to the modified Newtonian limit are that the
potential $\phi$, the function $\mathcal{N}$, and typical (squared)
velocities $\mathbf{v}^{2}$ are much smaller than unity; all terms
linear in these quantities are retained, but higher powers
($\phi^{2}$, $\cdots$) and products of them ($\phi \, \mathcal{N}$,
$\cdots$) are neglected. (In the application to galaxies this is an
excellent approximation.)  Apart from the rest energy density $\rho$
of the ordinary (``baryonic'') matter, the effective energy density
$\rho_{\text{eff}}$ contains the ``vacuum'' contribution
\begin{align}
\bigl( 8 \pi \, \overline{G} \, \bigr)^{-1}
\, \nabla^{2} \mathcal{N} (\mathbf{x}) =
\bigl( 8 \pi \, \overline{G}^{2} \, \bigr)^{-1}
\, \nabla^{2} G (\mathbf{x})
\label{45}
\end{align}
which is entirely due to the position dependence of Newton's constant.
Since it acts as a source for $\phi$ on exactly the same footing as
$\rho$ it mimics the presence of ``dark matter''.

As the density \eqref{45} itself contains a Laplacian $\nabla^{2}$,
all solutions of the Newtonian field equation \eqref{43} have a very
simple structure:
\begin{align}
\phi (\mathbf{x}) = \widehat \phi(\mathbf{x}) + 
\tfrac{1}{2} \, \mathcal{N} (\mathbf{x}).
\label{46}
\end{align}
Here $\widehat \phi$ is the solution to the standard Poisson equation
$\nabla^{2} \widehat \phi = 4 \pi \, \overline{G} \, \rho$ containing
only the ordinary matter density $\rho$. The simplicity and generality
of this result is quite striking.

Up to this point the discussion applies to an arbitrary prescribed
position dependence of Newton's constant, not necessarily related to a
RG trajectory. At least in the case of spherically symmetric systems
the identification of the relevant geometric cutoff is fairly
straightforward, $k \propto 1 / r$, so that we may consider the
function $G (k)$ as the primary input, implying $G (r) \equiv G ( k =
\xi / r)$. Writing again $G \equiv \overline{G} \, \left[ 1 +
  \mathcal{N} \right]$ we assume that $G (k)$ is such that 
$\mathcal{N} \ll 1$. Then, to leading order, the potential for a point
mass reads, according to \eqref{46}:
\begin{align}
\phi (r) = - \frac{\overline{G} \, M}{r}
+ \tfrac{1}{2} \, \mathcal{N} (r). 
\label{47}
\end{align}

Several comments are in order here.

\noindent
\textbf{(a) }The reader might have expected to find a term $-
\overline{G} M \, \mathcal{N} (r) / r$ on the RHS of \eqref{47}
resulting from Newton's potential $\phi_{\text{N}} \equiv -
\overline{G} M / r$ by the ``improvement'' $\overline{G} \to G (r)$.
However, this term $\phi_{\text{N}} \, \mathcal{N}$ is of second order
with respect to the small quantities we are expanding in. In the
envisaged application to galaxies, for example, $\phi_{\text{N}} \,
\mathcal{N}$ is completely negligible compared to the $\tfrac{1}{2} \,
\mathcal{N}$--term in \eqref{47}.

\noindent
\textbf{(b) }According to \eqref{47}, the renormalization effects
generate a nonclassical force (per unit test mass) given by
$-\mathcal{N}^{\prime} (r) / 2$ which adds to the classical $1 /
r^{2}$--term. This force is attractive if $G (r)$ is an increasing
function of $r$ and $G (k)$ a decreasing function of $k$. This is in
accord with the intuitive picture of the antiscreening character of
quantum gravity \cite{mr}: ``Bare'' masses get ``dressed'' by virtual
gravitons whose gravitating energy and momentum cannot be shielded and
lead to an additional gravitational pull on test masses therefore.

\noindent
\textbf{(c) }The solution \eqref{47} is not an approximation artifact.
In \cite{h2} we constructed exact solutions of the full nonlinear
modified Einstein equations (with $\mathcal{N}$ not necessarily small)
which imply \eqref{47} in their respective Newtonian regime. Those
exact solutions can be interpreted as a ``deformation'' of the
Schwarzschild metric ($M \neq 0$) or the Minkowski metric ($M=0$)
caused by the position dependence of $G$. The solutions related to the
Minkowski metric are particularly noteworthy. They contain no ordinary
matter (no point mass), but describe a curved spacetime, a kind of
gravitational ``soliton'' which owes its existence entirely to the
$\mathbf{x}$--dependence of $G$. At the level of eq.\ \eqref{47} they
correspond to the $M=0$--potential $\phi = \tfrac{1}{2} \,
\mathcal{N}$ which solves the modified Poisson equation if the
contribution $\propto \nabla^{2} \mathcal{N}$ is the only source term.
In the picture where dark matter is replaced with a running of $G$
this solution corresponds to a pure dark matter halo containing no
baryonic matter (yet). The fully relativistic $M=0$--solutions might
be important in the early stages of structure formation \cite{h2}.

Let us make a simple model of a spherically symmetric ``galaxy''. For
an arbitrary density profile $\rho = \rho (r)$ the solution of eq.\ 
\eqref{43} reads
\begin{align}
\phi (r) = \int \limits_{}^{r} \!\! \text{d} r^{\prime}~
\frac{\overline{G} \, \mathcal{M} (r^{\prime})}{{r^{\prime}}^{2}}
+ \tfrac{1}{2} \, \mathcal{N} (r)
\label{48}
\end{align}
where $\mathcal{M} (r) \equiv 4 \pi \, \int \limits_{0}^{r} \!\!
\text{d} r^{\prime}~ {r^{\prime}}^{2} \, \rho (r^{\prime})$ is the
mass of the ordinary matter contained in a ball of radius $r$. We are
interested in periodic, circular orbits of test particles in the
potential \eqref{48}. Their velocity is given by $v^{2} (r) = r \,
\phi^{\prime} (r)$ so that we obtain the rotation curve
\begin{align}
v^{2} (r) = \frac{\overline{G} \, \mathcal{M} (r)}{r} 
+ \frac{1}{2} \, r \, \frac{\text{d}}{\text{d} r} \, \mathcal{N} (r).
\label{49}
\end{align}

We identify $\rho$ with the density of the ordinary luminous matter
and model the luminous core of the galaxy by a ball of radius $r_{0}$.
The mass of the ordinary matter contained in the core is $\mathcal{M}
(r_{0}) \equiv \mathcal{M}_{0}$, the ``bare'' total mass of the
galaxy. Since, by assumption, $\rho=0$ and hence $\mathcal{M} (r) =
\mathcal{M}_{0}$ for $r > r_{0}$, the potential outside the core is
$\phi (r) = - \overline{G} \, \mathcal{M}_{0} / r + \mathcal{N} (r) /
2$. We refer to the region $r > r_{0}$ as the ``halo'' of the model
galaxy.

As an example, let us adopt the power law $G (k) \propto k^{-q}$ which
we motivated in Section \ref{s4}. We assume that this $k$--dependence
starts inside the core of the galaxy (at $r<r_{0}$) so that $G (r)
\propto r^{q}$ everywhere in the halo. For the modified Newtonian
limit to be realized, the position dependence of $G$ must be weak.
Therefore we shall tentatively assume that the exponent $q$ is very
small ($0 < q \ll 1$); applying the model to real galaxies this will
turn out to be the case actually. Thus, expanding to first order in $q$,
$r^{q} = 1 + q \, \ln (r) + \cdots$, we obtain $G (r) = \overline{G}
\, \bigl[ 1 + \mathcal{N} (r) \bigr]$ with
\begin{align}
\mathcal{N} (r) = q \, \ln (\kappa r)
\label{50}
\end{align}
where $\kappa$ is a constant. In principle the point $\overline{G}$
about which we linearize is arbitrary, but in the present context
$\overline{G} \equiv G_{\text{lab}}$ is the natural choice. In the
halo, eq.\ \eqref{50} leads to a logarithmic modification of Newton's
potential
\begin{align}
\phi (r) = - \frac{\overline{G} \, \mathcal{M}_{0}}{r} 
+ \frac{q}{2} \, \ln (\kappa r).
\label{51}
\end{align}
The corresponding rotation curve is
\begin{align}
v^{2} (r) = \frac{\overline{G} \, \mathcal{M}_{0}}{r} 
+ \frac{q}{2}.
\label{52}
\end{align}
Remarkably, at large distances $r \to \infty$ the velocity approaches
a constant $v_{\infty} = \sqrt{q/2\,}$. Obviously the rotation curve
implied by the $k^{-q}$--trajectory does indeed become flat at large
distances -- very much like those we observe in Nature.

Typical measured values of $v_{\infty}$ range from $100$ to
$\unit[300]{km/sec}$ so that, in units of the speed of light,
$v_{\infty} \approx 10^{-3}$. Thus, ignoring factors of order unity
for a first estimate, we find that the data require an exponent of the
order
\begin{align}
q \approx 10^{-6}.
\label{53}
\end{align}
The smallness of this number justifies the linearization with respect
to $\mathcal{N}$. It also implies that the variation of $G$ inside a
galaxy is extremely small. The relative variation of Newton's constant
from some $r_{1}$ to $r_{2}>r_{1}$ is $\Delta G / G = q \, \ln (r_{2}
/ r_{1})$. As the radial extension of a halo comprises only 2 or 3
orders of magnitude the variation between the inner and the outer
boundary of the halo is of the order $\Delta G / G \approx q$, i.\,e.\ 
Newton's constant changes by one part in a million only.

Including the core region, the complete rotation curve reads
\begin{align}
v^{2} (r) = \frac{\overline{G} \, \mathcal{M} (r)}{r} + \frac{q}{2}.
\label{54}
\end{align}
The $r$--dependence of this velocity is in qualitative agreement with
the observations. For realistic density profiles, $\mathcal{M} (r) /
r$ is an increasing function for $ r < r_{0}$, and it decays as
$\mathcal{M}_{0} / r$ for $r>r_{0}$. As a result, $v^{2} (r)$ rises
steeply at small $r$, then levels off, goes through a maximum at the
boundary of the core, and finally approaches the plateau from above.
Some galaxies indeed show a maximum after the steep rise, but
typically it is not very pronounced, or is not visible at all. The
prediction of \eqref{52} for the characteristic $r$--scale where the
plateau starts is $2 \, \overline{G} \, \mathcal{M}_{0} / q$; at this
radius the classical term $\overline{G} \, \mathcal{M}_{0} / r$ and the
nonclassical one, $q/2$, are exactly equal. With $q = 10^{-6}$ and
$\mathcal{M}_{0} = 10^{11} \, M_{\odot}$ one obtains $\unit[9]{kpc}$,
which is just the right order of magnitude.

The above $v^{2} (r)$ is identical to the one obtained from standard
Newtonian gravity by postulating dark matter with a density
$\rho_{\text{DM}} \propto 1 / r^{2}$. We see that if $G (k) \propto
k^{-q}$ with $q \approx 10^{-6}$ no dark matter is needed. The
resulting position dependence of $G$ leads to an effective density
$\rho_{\text{eff}} = \rho + q / \bigl( 8 \pi \, \overline{G} \, r^{2}
\bigr)$ where the $1/r^{2}$--term, well known to be the source of a
logarithmic potential, is present as an automatic consequence of the
RG improved gravitational dynamics.

We consider these results a very encouraging and promising indication
pointing in the direction that quantum gravitational renormalization
effects could be the origin of the plateaus in the observed galaxy
rotation curves. If so, the underlying RG trajectory of QEG is
characterized by an almost constant anomalous dimension
$\eta_{\text{N}} = - q \approx - 10^{-6}$ for $k$ in the range of
galactic scales.

Is the Einstein--Hilbert truncation sufficient to search for this
trajectory? Unfortunately the answer is no. According to eq.\ 
\eqref{16c}, $\eta_{\text{N}}$ is proportional to $g$ which is
extremely tiny in the regime of interest, smaller than its solar
system value $10^{-92}$. In order to achieve a $|\eta_{\text{N}}|$ as
large as $10^{-6}$, the smallness of $g$ must be compensated by large
IR enhancement factors. As a result, $\lambda$ should be extremely
close to $1/2$, in which case the RHS of \eqref{16c} is dominated by
the pole term: $\eta_{\text{N}} \approx - \left( 6 \, g / \pi \right)
\, \left( 1 - 2 \, \lambda\right)^{-1}$. Assuming $g \approx 10^{-92}$
as a rough estimate, a $q$--value of $10^{-6}$ would require $1 - 2 \,
\lambda \approx 10^{-86}$. It is clear that when $1 - 2 \, \lambda$ is
so small the Einstein--Hilbert trajectory is by far too close to its
termination point to be a reliable approximation of the true one.
Moreover, $\eta_{\text{N}} \bigl( g (k), \lambda (k) \bigr)$ is not
approximately $k$--independent in this regime. Thus we must conclude
that an improved truncation will be needed for an investigation of the
conjectured RG behavior at galactic scales, in particular to search
for the $g_{*}=0$--fixed point speculated about at the end of Section
\ref{s4}.

It is clear that the above model of a galaxy is still quite simplistic
and does not yet reproduce all phenomenological aspects of the mass,
size, and angular momentum dependence of the rotation curves for
different galactic systems. In particular $v_{\infty}$ is a universal
constant here and does not obey the empirical Tully--Fisher relation.
Similarly, according to MOND, the need for a dark matter substitute
actually does not arise at a critical distance but at a critical
acceleration. As we explained in \cite{h2} to which the reader is
referred for further details these limitations are due to the
calculational scheme used here (``cutoff identification'', etc.).
Usually this scheme can provide a first qualitative or
semi--quantitative understanding, but if one wants to go beyond this
first approximation, a full fledged calculation of $\Gamma [g_{\mu
  \nu}]$ would be necessary which is well beyond our present technical
possibilities.
%
%
%
%
%
%
%
%
\section{Summary and conclusion}
\label{s6}
In this paper we assumed that Quantum Einstein Gravity correctly
describes gravity on all length scales and tried to identify the
underlying RG trajectory realized in Nature. Along this trajectory, we
found a regime where the running of the gravitational parameters is
unmeasurably small, the domain of classical General Relativity. The
renormalization effects become strong both at momentum scales $k$ much
larger (UV regime) and much smaller (IR regime) than those of the GR
regime. In the UV regime, for $k \to \infty$, the trajectory
approaches a non--Gaussian fixed point. By definition, QEG is the
``asymptotically safe'' theory whose bare action equals the fixed
point action, guaranteeing its nonperturbative renormalizability. We
analyzed a potential IR divergence (instability) in the flow equation
which occurs in the Einstein--Hilbert truncation already, and we
argued that it triggers significant renormalization effects in the
IR, i.\,e.\ at distances larger than those where classical GR was
successfully tested. In particular, Newton's constant increases for
decreasing $k$ so that $G$ should appear to be larger at galactic or
cosmological scales than in a terrestrial or solar system
``laboratory''. Looking for possible manifestations of this effect we
analyzed the possibility that the almost flat plateaus observed in the
rotation curves of spiral galaxies, usually attributed to the presence
of dark matter, are actually due to the RG running of Newton's
constant. While the galactic regime is inaccessible with the
Einstein--Hilbert truncation, a phenomenological analysis in the
framework of the improved action approach revealed that a power
law--type scale dependence $G (k) \propto k^{-q}$, corresponding to an
(approximately) constant anomalous dimension $\eta_{\text{N}}$, leads
to rotation curves which are in qualitative agreement with those found
in Nature. In particular they become perfectly flat at sufficiently
large distances.

As a by--product we arrived at a new understanding of the cosmological
constant problem: By analyzing the RG trajectories among which Nature
could have chosen we saw that there exist no trajectories which, at
the same time, would predict a long classical regime and a large
cosmological constant. If there is a Universe inhabited by human
beings who manage to verify that gravity behaves according to
classical General Relativity over many different length scales, they
unavoidably will observe a cosmological constant which is extremely
small compared to the value of $1/G \equiv m_{\text{Pl}}^{2}$ they
measure. Conversely, if $\Lambda$ is not small, it is also not
constant, so GR is not valid, and there is no point wondering about
the value of a non--constant ``constant''.  This argument does not
seem to require a detailed understanding of the IR effects since the
data indicate that in our world, between the end of the GR regime and
the Hubble scale, $G$ and $\Lambda$ changed by comparatively small
factors only.

As for the IR effects, it will be exciting to see whether the
properties of the RG trajectory of QEG which we predicted can be
confirmed by more advanced ab initio calculations. Clearly more
general truncations for the effective average action, involving
nonlocal invariants for instance, should be explored for this purpose.
In view of recent progress \cite{ajl} made on causal (Lorentzian)
dynamical triangulations \cite{amb} one also may hope that at some
point it will be possible to make contact with numerical simulations.
This could lead to an independent confirmation of the picture we have
drawn in the present paper.

\vspace*{3.3mm}
\noindent 
Acknowledgement: We would like to thank A.~Bonanno, O.~Lauscher,
R.~Loll, F.~Saueressig, and J.~Smit for helpful discussions.
%
%
%
%
%
%
%
%

\end{document}